\documentclass[]{emulateapj-rtx4}

\usepackage{times}
\usepackage{epsfig}

\usepackage{longtable}
\usepackage{rotating}

\DeclareMathAlphabet{\mathsc}{OT1}{cmr}{m}{sc}
\def\testbx{bx}%
\DeclareRobustCommand{\ion}[2]{%
\relax\ifmmode
\ifx\testbx\f@series
{\mathbf{#1\,\mathsc{#2}}}\else
{\mathrm{#1\,\mathsc{#2}}}\fi
\else\textup{#1\,{\mdseries\textsc{#2}}}%
\fi}

\newcommand{\rsun}{\mbox{\,$\rm R_{\odot}$}}        

\shorttitle{Coronal lines}
\shortauthors{Del Zanna et al.}

\begin{document}

\title{Solar coronal lines in the visible and infrared. A rough guide}

\author{Giulio Del Zanna}
\affil{DAMTP, CMS, University of Cambridge, Wilberforce Road, Cambridge CB3 0WA, United Kingdom}
 \email{gd232@cam.ac.uk}

\author{Edward E.  DeLuca}
\affil{Harvard-Smithsonian Center for Astrophysics, 60 Garden Street, Cambridge, MA 02138, United States}

\begin{abstract} 
We review the coronal visible and infrared lines, collecting  previous observations,
and comparing, whenever available, observed radiances with those predicted by 
various models: the quiet Sun, a moderately active Sun, and an active region as
observed near the limb, around 1.1\rsun. We also model the off-limb radiances 
for the quiet Sun case. We used the most up-to-date atomic data in CHIANTI
version 8.
The comparison is satisfactory, in that all of the strong 
visible lines now have a firm identification. We revise several previous identifications
and suggest some new ones. 
We also list the large number of observed lines for which we do not currently
have atomic data, and therefore still await firm identifications.
We also show that a significant number of coronal lines should be observable
in the near-infrared region of the spectrum by the  upcoming Daniel K. Inouye Solar Telescope
(DKIST) and the  AIR-Spec instrument,
which  observed  the corona during the 2017 August 21 solar eclipse. 
We also briefly discuss the many potential spectroscopic diagnostics available
to the visible and infrared, with particular emphasis on measurements of 
electron densities and chemical abundances. We briefly point out some of the potential 
diagnostics that could be available with the future infrared instrumentation that is being built
for DKIST  and planned for the 
 Coronal Solar Magnetism Observatory (COSMO).
Finally, we highlight the need for further improvements in the atomic data.
\end{abstract}


\section{Introduction}

Within the solar community, 
there is currently renewed interest in the coronal forbidden  lines in the visible and infrared,
partly because  several breakthrough observations are forthcoming.
For example, the Daniel K. Inouye Solar Telescope
 (DKIST, see \citealt{rimmele_etal:2015}),  formerly known as ATST \citep{keil_etal:2003},
will carry out ground-breaking observations of the 
solar corona with its 4 meter telescope on  Haleakala, Maui 
with two dedicated  coronagraphs.
It will begin operations in 2019 and will carry out routine daily 
observations up to 1.5\rsun\  in selected wavelength regions
between about 5000~\AA\ and 5~$\mu$m with  the CryoNIRSP 
spectropolarimeter \citep{fehlmann_etal:2016}. In its coronagraphic mode,
the instrument will have a resolution of 30,000, a spatial resolution of 0.5\arcsec, 
and a field of view (FOV) of 4\arcmin\ $\times$3\arcmin, equivalent to the size of an 
active region. The FOV will be scanned with high cadence using a multislit. 
In addition, global scale coronal magnetic field measurements are the focus
 of the proposed Coronal Solar Magnetism Observatory (COSMO),
see \cite{tomczyk_etal:2016}.

The longer wavelengths of the infrared spectral region are still largely unexplored,
but there is a lot of interest in this spectral region, and in the near future
several new observations, aside from DKIST, will become available.
 For example, the 
airborne infrared spectrometer (AIR-Spec), described by \cite{deluca_etal:2016},
 was built to  carry out observations in the  infrared  during  the 2017 August 21 solar eclipse, 
on-board an high-altitude airplane. The instrument was designed to observe four spectral 
regions (2.82--3.07~$\mu$m and 3.74--3.98~$\mu$m in first order),
and five coronal lines: \ion{Si}{x}  1.43~$\mu$m,  \ion{S}{xi}  1.92~$\mu$m,
 \ion{Fe}{ix}  2.86~$\mu$m,
\ion{Mg}{viii}  3.03~$\mu$m, and \ion{Si}{ix}  3.93~$\mu$m. 
Preliminary analysis (Samra 2017 private communication)
suggests that all of the lines were successfully observed.

One might ask: what could  observations of the 
coronal visible and infrared lines provide, given the abundance of 
current EUV and UV observations?

The importance of the infrared observations for measuring
the coronal magnetic field has been recognised for a long time,
see e.g. the \textit{Living Review} by  \cite{penn:2014}.
\cite{judge:1998} carried out  simulations
to predict which of the  visible and infrared coronal lines, 
should be observable, 
with an emphasis on the infrared ones which could be
 used  to measure magnetic fields.
\cite{judge_etal:2013} discussed some specific issues related to 
DKIST.

The forbidden lines are also useful for many other diagnostics,
aside from the potential to measure coronal magnetic fields.
For example, plasma flows and line widths can be measured with an accuracy 
far superior to that of  any other wavelength.
There is an extended literature on the non-thermal widths of the lines, which
in principle provide important constraints to coronal heating theories.

Another example concerns the possibility to  provide direct measurements of electron 
temperatures  and constrain for the presence of 
non-Maxwellian electron distributions \citep{dudik_etal:2014_fe}, 
when combined with measurements in the UV or EUV.
Such diagnostics have been attempted in the EUV 
\citep{dudik_etal:2015_loop}, but are very difficult.

For a general review on the diagnostic 
potential of the forbidden lines  from ground-based observations
see   \cite{landi_etal:2016}. 
In their  review, future ground-based instrumentation is also
briefly presented, with emphasis on UCoMP and COSMO.
UCoMP is an upgraded version of the 
 Coronal Multichannel Polarimeter (CoMP) instrument, 
described by \cite{tomczyk_etal:2008}.
UCoMP will be  a pathfinder for a Large 1.5 m refractive Coronagraph
(LC), which is  currently in its advanced planning phase,  and forms part of a 
suite of ground-based telescopes, COSMO, which will be built to 
study the outer corona.
The current design of the COSMO LC calls for a tunable filter focal plane 
instrument operating between 500-1100nm. 
The baseline covers cool CME core lines (\ion{H}{i}, \ion{He}{i},
 \ion{Ca}{ii}, \ion{O}{ii}, \ion{O}{iii}, and  \ion{Fe}{iv}), 
hot lines  (\ion{Fe}{xiv}, \ion{Fe}{xv}, \ion{S}{xii}, 
\ion{Ar}{xiii} and \ion{Ca}{xv}) and lines in the quiescent corona 
(\ion{Fe}{x}, \ion{Fe}{xi}, \ion{Fe}{xiii}, \ion{Fe}{xiv},
 \ion{Ar}{x}, and \ion{Ar}{xi}). 
The spatial resolution of COSMO LC is 2" and the FOV covers a full degree.
 We can see that it will be an excellent complement to DKIST 
providing a high magnetic sensitivity (1G), temperature coverage and global coronal coverage.

In the present paper, we limit our 
discussion to  diagnostic techniques of line
intensity ratios  which can be used to measure 
the electron density and relative elemental abundances.

As already pointed out by \cite{woolley_allen:1948},
the visible lines allow a direct measurement of the 
absolute (i.e. relative to hydrogen) values of the chemical abundances, 
by comparing line radiances with the continuum. 

It is well-known  that coronal abundances can be different 
from the photospheric ones, and that correlations
exist between the  abundance of an element   and its 
first ionization potential (FIP).
The low-FIP ($\leq$ 10 eV) elements are more abundant 
than the high-FIP ones, relative to their  photospheric values.
See, e.g. the \textit{Living Review} by \cite{laming:2015}.
The FIP effect is now considered an important issue, because it
can help to trace the in-situ solar wind measurements 
to the source regions, an important topic for 
the upcoming Solar Orbiter (to be launched in early 2019), 
which will measure elemental abundances with both in-situ and
remote-sensing instruments.
Relative abundances are comparatively easier  to measure and there are plenty
of observations in the X-ray, EUV, and UV. Absolute 
values are harder to obtain, and in the literature
there are plenty of  contradictory results. In fact, it is not
well-accepted if low-FIP elements are over-abundant, or the 
high-FIP ones  are depleted, compared to the photospheric values. 
 There are  SoHO SUMER and UVCS measurements
relative to hydrogen lines, but a direct measurement against the 
continuum is only available in the X-rays during flares.
Therefore, line-to-continuum measurements of the forbidden lines
in the visible and infrared have a potential to resolve these issues,
although we should point out that an instrument such as 
 DKIST will only measure the polarized component of the
continuum, and not the total continuum.

Aside from all the above diagnostic possibilities, the forbidden lines
are also very useful just to trace the open and closed structures in the 
outer corona.
The forbidden  lines are very sensitive to photoexcitation 
of the disk brightness, at the low densities of the outer corona.
Therefore, their intensities are closer to being proportional to the electron density,
rather than the square,  as  is the case for the strong 
dipole-allowed lines in the EUV.
This means that the intensities of the forbidden lines  decrease with the radial 
distance more slowly than those of the EUV lines, hence 
they show the  outer corona  to larger distances.

Comparisons of coronal emission lines in the EUV and Vis/IR show enhanced
radiative contributions in the visible/IR forbidden lines 
\citep{habbal_etal:2010,habbal_etal:2011}. 
 Habbal's work used narrow band filters
centered on and off strong emission lines, allowing the continuum
contribution to be isolated from the line contribution. Narrow band imaging
allows for the exploration of the whole corona during each eclipse
observation. The success of her imaging approach motivates our need for an
accurate understanding of the visible/IR coronal emission.

It might be surprising, but the visible and infrared
regions  of the coronal spectrum are largely unexplored. 
This is mainly due to  the paucity of observations. 
Interestingly, the same problem concerns the soft X-rays,
between 60 and 170~\AA. The two spectral regions are
historically related: Edl{\'e}n's pioneering laboratory
measurements of  soft X-ray spectra in the 1930s and 
identifications of  lines from iron and other elements 
allowed \cite{grotrian:39}  to identify  the famous 
bright red coronal forbidden line at 6374.6~\AA\
as the \ion{Fe}{x} transition
$^2$P$_{3/2}$--$^2$P$_{1/2}$  within the ground configuration. 
 Edl{\' e}n  then wrote a seminal paper \citep{edlen:43},
confirming that the solar corona is a million degree plasma
(see also \citealt{swings:43} for more details).
Edl{\'e}n work was extended by \cite{fawcett_etal:72}, and
recently revised by \cite{delzanna:12_sxr1}, where the strongest
lines in the soft X-ray spectra, which were previously unknown, have been identified.
This was only possible because of large-scale atomic calculations for the main 
iron and nickel coronal ions which were recently carried out
within the UK APAP network\footnote{\url{http://www.apap-network.org}},
 the main provider of atomic data for fusion and astrophysical plasma.
However, a large number of lines in the soft X-rays are still unidentified.

In the literature, contradicting line identifications of the visible lines
are common, even for some among the strongest lines. 
A large number of authors have contributed to the 
identifications over the years. Among those:
Edl{\' e}n, Svensson, Ekberg, Smitt,  Mason,  Nussbaumer, Magnant-Crifo,
just to name a few. 
 Most of the suggested identifications 
were mainly based on wavelength coincidences along  isoelectronic 
sequences, and sometimes with approximate estimates of the line intensities.
 One exception among the early studies is the work of \cite{mason_nussbaumer:77},
where the excitation cross section  for 
\ion{Fe}{x} and  \ion{Fe}{xi} were actually calculated, 
allowing more accurate estimates of line intensities. This enabled 
the authors to suggest, also on intensity grounds, 
several new identifications for these two important ions.

It might be surprising, but  accurate atomic rates 
for the forbidden lines have only been recently available for a number of ions.
The main ions producing the strongest forbidden lines are from Iron,
and the above-mentioned large-scale atomic calculations have produced
significantly different results. 
For example, as shown in \cite{delzanna_etal:2014_fe_9}, significant differences (factors of 2)
were found for some of the forbidden lines of \ion{Fe}{ix} and
\ion{Fe}{x}, compared to earlier (but still sophisticated) calculations.
Most of these data have been included in 
CHIANTI\footnote{\url{http://www.chiantidatabase.org/}} 
version 8 \citep{delzanna_chianti_v8}, which is used here.
We note that 
\cite{judge:1998} used the atomic data available at the time, i.e.
CHIANTI version 1 \citep{dere_etal:97}.

It is therefore timely to review all the  observations of the forbidden lines
with the latest atomic data.
Our main aim of this paper is to provide a comprehensive list
 of the main lines observed, and assess their identification on the basis 
of the current atomic data. 
A full benchmark of the atomic data along the lines of the 
series of papers which  started with \cite{delzanna_etal:04_fe_10}
 is not possible at this time,
given the uncertainties in the calibrated radiances
of the visible and infrared lines.
Many identifications will need to be confirmed with future observations.

Our aim is also  to provide rough estimates of line radiances
for different activity conditions, and as a function of distance,
to aid the planning of future observations.
 In particular, the present results were used 
to aid the planning of the AIR-Spec  infrared observations. 
Finally, we also provide some examples of 
which spectral lines would be  suitable for measuring 
electron densities and  elemental abundances, using intensity
ratios.

\section{Observations}

In this section, we briefly review only the main observations we have used 
for the present assessment, and which contained  published 
spectral observations where wavelengths and/or  line radiances were provided.
In other words, this section is not meant to even summarise the 
large volume of literature on ground- and space-based observations 
of the forbidden lines, especially during total eclipses.

\subsection{Observations in the visible}

The genius of Bernard Lyot  made breakthrough observations of the 
coronal visible and near infrared lines with a novel coronagraph
outside eclipses. He carried out observations at the Pic du Midi
Observatory, observing 11 coronal lines, some for the first time
\citep{lyot:1939}. In many respects, his observations are still 
unsurpassed.
Lyot also  classified the lines into three groups, following
their intensity distributions in the corona. 
The three representatives of the groups were the 
red \ion{Fe}{x}, the green  \ion{Fe}{xiv}, and the yellow \ion{Ca}{xv}.
We now know that this simple classification reflects the 
temperature of formation of these ions, and it is a very useful information 
when trying to identify a spectral line. 
This scheme was somewhat complicated later on by 
\cite{shajn:1948} and following authors with the introduction of 
a fourth class. So after \cite{vandehulst:1953}, it has been
customary to assign lines from ions such as \ion{Fe}{x} a class I,
those like \ion{Fe}{xiii} and \ion{Fe}{xiv} a class II,
the \ion{Ca}{xiii} and \ion{Ni}{xvi} a class III, while all the 
hotter ions such as \ion{Ca}{xv} and  \ion{Fe}{xv},
 only visible in active regions, were given a class IV.

We note that in the literature it was common to name 
the bright hot regions in the solar corona as `condensations'. 
In the present paper, we
 refer  to them as active regions, because 
that is what they were. 
We also note that we only consider here coronal lines
formed around 1 MK or more, and not chromospheric lines.

There are many historical observations of coronal lines during 
eclipses. An early one worth a special mention is 
 the  Harvard-MIT expedition to the 1936 eclipse.
The expedition  recorded spectra in the 
near-ultraviolet and visible with three instruments.
\cite{petrie_menzel:1942} provided an extensive list of lines, some  of which have not been 
observed since, hence should be considered with caution. 
We only report the coronal lines that they considered more certain.
A complete review list of 27 well-observed  coronal lines known 
at the time was provided by  \cite{vandehulst:1953}.

We now briefly review the main observations we have assessed, starting with the
two most important eclipses.
The present study extends  the excellent list 
of observed visible lines compiled by \cite{mouradian:1997},
by adding other observations.
It is also an improvement in that several of the identifications
listed by \cite{mouradian:1997} turned out to be incorrect.

\subsubsection{The 1952 February 25 eclipse }

Lyot designed a novel instrument with a circular 
slit, and built two spectrographs, one in the visible and one in 
the near ultraviolet, to observe the 1952 February 25 eclipse
at Khartoum, Egypt, with the assistance of Aly. They obtained 
the best ever plates of the coronal lines. They were also
lucky in that a very extended, hot and high-density 
active region was located at the limb. Next to it, a 
large prominence was also present, which was key to 
obtain accurate measurements of the coronal lines using as a 
reference the known  chromospheric lines, which were very strong 
in the prominence.

Unfortunately,  Lyot only managed to perform an initial 
analysis and died in Egypt three months after. 
The results of this initial analysis are reported in 
\cite{lyot_aly:1955} as a first Table. Aly subsequently 
added several weaker lines, and added a second Table
in the same publication. 
A preliminary photometric analysis of the plates was  later
carried out in  Meudon by \cite{divan_pecker:1960},
while a further analysis and line intensities were provided in
\cite{aly_etal:1962}.
We adopt in our Table~\ref{tab:big_table}  the calibrated radiances at the 
center of the active region, as listed by \cite{jefferies_etal:1969} 
(see below). 
As \cite{aly_etal:1962} described, the relative  calibration was
quite accurate, because it could be checked against the observed 
continuum.  However, the absolute values are uncertain 
by about a factor of two.
The distance from the limb was also quite uncertain, but should range
between 1.05 and 1.1\rsun, with 1.09\rsun as the quoted value.

The plates were subsequently taken to Sac Peak and re-analysed by Aly,
Evans and Orrall independently, as described in \cite{aly_report}.
The lines were given a class, on the basis of the certainty in the 
identification of a line as a coronal line in the plates, which was
not trivial. 
The well-known lines were given an 'A'.
The well-observed ones were given a class 'B', while those more
difficult to measure were listed as 'C'. The dubious ones
were given a class 'D'.
Several of the lines originally listed by \cite{aly_etal:1962}
in the second table were subsequently listed as dubious by \cite{aly_report}.
In our  Tables~\ref{tab:big_table},\ref{tab:obs_not} below we
typically list only the lines in the A,B,C classes
as in the \cite{aly_report} report.

The wavelengths are often given with a decimal figure, but appear to be 
 uncertain by about 1--3~\AA,  as it appears from a comparison with subsequent 
observations. This means that in several instances it is not 
clear if more lines have been observed or just the 
wavelengths reported were inaccurate. 
When the differences are more than  1~\AA\ we list
the wavelengths reported by Aly.

Lyot's observations had a profound influence in many areas.
They showed that it is possible to 
measure the absolute abundances (i.e. relative to 
hydrogen) of the observed elements. This was carried out by 
\cite{pottasch:1964} with these data, and later by several 
authors with these and other datasets \citep[see, e.g.][]{magnant-crifo:1974,mason:1975}

Lyot's  observations prompted  M.J. Seaton  at University College London
(UCL)  to challenge  H.E. Mason to provide the first accurate 
scattering calculations for the coronal iron ions \citep{mason:1975a}.
The calculations  used the  state-of-the-art 
UCL \emph{Distorted Wave} (DW) scattering codes,
mainly developed by W. Eissner and M.J. Seaton
\citep{eissner_seaton:1972}. 
These calculations  allowed for 
the first time one to predict line intensities.
It was shown that cascading is an important effect when 
estimating the intensities of the forbidden lines.
 These codes
have been widely used since for the calculations of atomic data 
in general for astrophysics.


\subsubsection{The 1965 May 30 eclipse}

Two spectrographs similar to the one used by Lyot 
(with a  circular slit)
were built by the Sac Peak and University of Hawaii groups
under the supervision of Dunn \citep{dunn:1966}, to observe the 1965 May 30 eclipse.
One spectrograph was aboard a NASA aircraft and 
recorded spectra in the visible from 3000 to 9000~\AA. 
The other spectrograph obtained ground-based simultaneous observations
from the Bellingshauen Island.

The spectra were dominated by the presence of an active region,
which however was not as bright and hot as that one observed by Lyot 
in 1952.
\cite{curtis_etal:1965} presented a complete list of wavelengths
and identifications based on the flight spectra.
We adopt it as the basis of our Tables~\ref{tab:big_table},\ref{tab:obs_not}
for the lines in the 3000 to 9000~\AA\ range. 
The estimated accuracy is about 0.4~\AA.
We have converted the  air  wavelengths measured by Curtis
 to vacuum wavelengths following \cite{edlen:66}.
The Lyot class of the observed lines is also 
reported in the  Tables, whenever available.

\cite{jefferies_etal:1971} provided a well-cited list 
of line radiances obtained from the ground at two slit positions.
The wavelengths listed by \cite{jefferies_etal:1971} are those 
reported by \cite{curtis_etal:1965}.

As discussed by 
\cite{jefferies_etal:1969}, the same instrument obtained 
good spectra at the 1966 eclipse in Bolivia. 
\cite{jefferies_etal:1969} provides a list of calibrated radiances of the 
1965 and and 1966 eclipse, as well as radiances in the core of the 
active region of the 1952 eclipse. 
This comparison is really important, as it shows that the 
AR observed by Lyot in 1952 was unusually bright and hot.
Indeed lines such as \ion{Ca}{xv} were not observed in the 
1965 and 1966 eclipses.

Several identifications listed by \cite{curtis_etal:1965}
and \cite{jefferies_etal:1971} turn out to be incorrect,
as discussed by various authors later on.
In most cases, lines were actually due to Iron ions. 
Most of the correct identifications are reported in the review 
by \cite{smitt:1977}, although the actual original
identifications are due to  many authors.

\cite{magnant-crifo:1973} later performed a detailed analysis 
of the flight spectra of the 1965 eclipse. They were taken 
at radial distances of about 1.1~\rsun.
Discrepancies with the 
radiances  obtained 
from the ground spectra were ascribed to an incorrect measurement
of the radial distance of the slit by \cite{jefferies_etal:1971}.
The useful result of the \cite{magnant-crifo:1973} analysis 
is a complete list of line equivalent widths as a function of positon 
angle and radial distance, from which line intensities can be obtained.
We have analysed the intensities at several positions, and noted that 
an active region dominated the spectra.
We selected two regions far away to compare with our estimates
(see below).

\subsubsection{Other observations}

The most remarkable observation of the visible forbidden lines in the 
red and near-infrared was obtained by
\cite{wlerick_fehrenbach:1963} with the 1.9 meter telescope of 
Haute provence, which happened to be along the totality path 
of the famous 1961 February 15 eclipse in Europe. 
They observed a number of  lines never observed since,
and provided the best ever wavelength measurements.
However, many lines were very weak so we list them with a question mark.

Many other observations have been carried out. We only 
briefly mention the most notable ones.
\cite{kurt:1962} observed the visible spectrum from an aircraft,
during the same eclipse of  1961 February 15.
\cite{kernoa_servajean:1965} report an analysis of other observations
taken during the same eclipse.
\cite{byard_kissel:1971} obtained the Fe abundance from observations
of the infrared Fe XIII lines during the total eclipse of 12 Nov 1966.
\cite{nikolsky_etal:1971} report several new weak unidentified lines 
observed during the 1970 March 7 eclipse. The wavelengths are not 
very accurate and most of the weak lines have not been observed 
by others, so are not reported here.
\cite{liebenberg_etal:1975} carried out detailed measurements of the 
coronal green line.

\subsection{Infrared observations}

Following the observations of the \ion{Fe}{xiii} infrared lines 
by \cite{lyot:1939},  many  observations of these and other lines have been carried out 
in a number of sites, see e.g. 
\cite{firor_zirin:1962,zirin:1970,fisher_pope:1971,querfeld:1977,arnaud_newkirk:1987,singh_etal:2004}.
More recently, routine observations of the \ion{Fe}{xiii} infrared lines 
 have been carried out with the  CoMP instrument \citep{tomczyk_etal:2008} at
 the Mauna Loa Solar Observatory.

Lines at longer wavelength are intrinsically weaker and
difficult to observe from the ground.
\cite{munch_etal:1967} reported  observations in the infrared, obtained with a
spectrograph and a circular slit close to the solar limb, obtained
during the 1966 November 12 eclipse from a NASA airborne flight.
The Si X 1.43 micron line was observed, as well as the 
Mg VIII 3.027 micron line. Approximate radiances for these lines were provided.
Note that the Si X 1.43 micron line was later also 
observed  with a 
coronagraph of the National Solar Observatory at Sacramento Peak by 
\cite{penn_kuhn:1994}.

\cite{olsen_etal:1971} carried out some excellent observations
in the infrared between 1 and 3~$\mu$m, obtained during the eclipse of 1970 March 7.
The spectrograph was mounted 
 on-board a high-altitude airplane. Approximate line intensities were
provided, as well as several identifications. 
We list  below in our Tables~\ref{tab:big_table},\ref{tab:obs_not} their measurements, but note that 
some of the identifications were incorrect, as pointed 
out later by \cite{kastner:1993}.

\cite{kuhn_etal:1996} presented an infrared spectrum 
  obtained during the 1994 solar eclipse. 
The S IX line at 1.25~$\mu$m was detected, and the 
\ion{Si}{x} 1.43~$\mu$m was also observed.
Finally, we note that \cite{judge_etal:2002} detected the \ion{Si}{ix} 3.93~$\mu$m line
from ground-based observations.

\section{Estimates of line radiances near the limb}

The Sun is a variable star, and it has always been known 
even from the early eclipse observations that the solar corona
presents completely different characteristics in time and space.
To provide a guideline of observable lines and assess the 
identifications of all the observed lines is a difficult task.

The forbidden lines are extremely sensitive to the electron temperature,
density, photoexcitation and chemical abundances. 
All these parameters change significantly from region to region.

We now have excellent spectroscopic measurements of the inner 
corona, up to 1.3\rsun, from e.g. SoHO CDS and SUMER,
Hinode EIS, and of the outer corona from SoHO UVCS.
We therefore have a fairly good understanding of these 
parameters to be able now to perform some reliable estimates
for the forbidden lines.

The easiest case to model is that of the quiet Sun (QS),
when no active regions are present.
However, very few measurements are available for the 
coronal lines, as most of the reported  observations are on 
active regions. 

In order to provide a rough guide on what radiances are expected
when the Sun is more active, we have therefore 
opted to present the results of two other simulations,
one on the brightest active region observed so far (AR) by Lyot
in 1952, and one  when the Sun was more quiet  (QR). 

As the forbidden lines are well known to be very sensitive to
the photoexcitation, we started by building a photospheric 
spectrum to be used for the simulations. 
We opted for the irradiance spectrum compiled by \cite{kurucz:2006},
which was based on high-resolution measurements, scaled 
to agree with the well-calibrated ATLAS-3 spectrum by 
\cite{thuillier_etal:2004}.
We supplemented this spectrum with that of a black body 
at the near-infrared wavelengths. 
We compared the \cite{kurucz:2006}  spectrum to that of a black-body 
at 6,000 K and to the lower-resolution reference spectrum
compiled by \cite{woods_etal:2009_whi}, and found some differences,
although not too significant for the present purpose.

We assume uniform distribution on disk, i.e.
have converted the irradiances to radiances. 
This is clearly a good assumption for the quiet Sun,
but not in the case of the active Sun. 
We have performed the simulations with and without
photoexcitation, to assess which line is most affected. 
We used the framework developed for CHIANTI version 
4  \citep[][]{young_etal:2003} by P.R. Young to include the
photoexcitation in the modelling.

\subsection{The QS simulation}

SoHO SUMER (see, e.g. \citealt{feldman_etal:97,landi_etal:2002})
 and Hinode EIS \citep[see, e.g.][]{warren_brooks:2009, delzanna:12_atlas} 
observations of the off-limb quiet Sun 
have established that the emission measure (EM) distributions
are  very narrow around 1 MK. 
This is obtained with the EM loci method \citep[see, e.g.][]{strong:78,delzanna_etal:02_aumic}, 
 by  plotting the ratio 
of the observed intensity $I_{\rm ob}$ of a line with its 
contribution function,
$I_{\rm ob} / G(T)$,  as a function of temperature.   The loci of these curves 
are an upper limit to the  emission measure distribution.
If the curves cross at one point, that is a strong indication of 
an isothermal plasma.

There is also good agreement in the absolute values, 
if one compares e.g. the EM loci curves from EIS and SUMER
\citep{delzanna:12_atlas,landi_etal:2002}.
We have chosen  as a  baseline for our model the SUMER observations
reported by  \cite{landi_etal:2002}, because they 
include far more ions than the EIS ones.
The SUMER spectra were taken between 1.03 and 1.045\rsun.
It is well-known from SoHO CDS, SUMER and Hinode EIS observations
of line ratios that the electron density is about 
2$\times 10^8$ (cm$^{-3}$), which is what we adopted. 
We run a simple 
isothermal model with 
log $T$ [K]=6.14,  log EM=27.1, 
and the photospheric abundances of \cite{asplund_etal:09}.
The observed and predicted SUMER radiances for a selection of 
strong lines  are shown in  Table~\ref{tab:sumer}.

What is not well established in the literature is the chemical 
composition of the low corona of the quiet Sun.
In most literature, including \cite{landi_etal:2002}, it is assumed that 
in the quiet corona the low-FIP elements are increased 
by a factor of about 3-4, compared to their photospheric values.
However, there is also a body of literature 
showing no FIP effect in the quiet Sun. 

For the purpose of the present estimates, this issue is not
relevant, as long as the model is capable of reproducing 
the observed line intensities.
However, it is interesting to note, looking at 
 Table~\ref{tab:sumer}, the relatively good agreement for both 
low- and high-FIP elements. We originally aimed at a factor of two agreement,
but in many instances the agreement is far better.
We also note that we selected 
those lines where the atomic data are more reliable.
In fact, many of the SUMER lines are forbidden and their intensities
are more difficult  to calculate.
Indeed, for many of the ions observed by SUMER, we still do not have 
accurate atomic data for all transitions.

\begin{table}[!ht]
\begin{center}
\caption{SUMER quiet Sun radiances  (erg cm$^{-2}$ s$^{-1}$  sr$^{-1}$)
of a selection of lines, as observed (I$_{\rm obs}$) by 
 \cite{landi_etal:2002} and predicted (I$_{\rm pred}$) with 
an isothermal assumption and the photospheric abundances of \cite{asplund_etal:09}.
\label{tab:sumer}
}
\begin{tabular}{llllllll}
\tableline
\tableline
  Ion &   $\lambda$ (\AA)  &  I$_{\rm obs}$ &  I$_{\rm pred}$ &  \\ 
\tableline
\noalign{\smallskip}

\ion{Fe}{x} & 1463.49  &  0.84 &  0.76 &   \\

\ion{Fe}{xi} & 1467.07  & 3.4  &  3.1 &   \\

\ion{Fe}{xii} & 1349.40  & 5.3  &  3.4 &   \\

\ion{Mg}{viii} & 772.26  & 0.67  &  0.42 &   \\
\ion{Mg}{ix} & 706.06  &  8.16 &  6.0 &   \\
\ion{Mg}{x} & 609.79  & 1.5$\times$10$^{2}$  &  1.4$\times$10$^{2}$ &   \\

\ion{Si}{vii} & 1049.22  & 0.04 & 0.03 & \\
\ion{Si}{viii} & 944.47  & 2.0  &  1.7 &   \\
\ion{Si}{ix} & 950.16  &  1.9 &  1.0 &   \\
\ion{Si}{x} & 638.93  & 5.2  &  3.9 &   \\
\ion{Si}{xii} & 520.67  & 9.2  &  10. &   \\

\ion{Ca}{x} & 557.77  &  8.0 &  10. &   \\
\noalign{\smallskip}
\tableline
\noalign{\smallskip}
\multicolumn{5}{c}{High-FIP elements} \\
\noalign{\smallskip}
\ion{Ne}{vii} & 895.17  &  0.13 &  0.18 &   \\
\ion{Ne}{viii} & 770.43  & 30.7  &  50. &   \\
\ion{O}{vi} & 1037.61  & 28.5  &  26. &   \\

\ion{S}{ix} & 871.73  & 0.44  &  0.29 &   \\
\ion{S}{x} & 776.37  &  1.9 &  1.6 &   \\

\ion{S}{xi} & 575.0  &  0.47  &  0.37 &   \\

\ion{Ar}{xi} & 1392.10  & 0.13  &  0.11 &   \\
\noalign{\smallskip}
\tableline
\tableline
\end{tabular}
\end{center}
\end{table}

We used this simple isothermal model to predict the radiances
of all the forbidden lines in the visible and infrared. 
In order to make a meaningful comparison to the eclipse data
taken at 1.1\rsun\  by \cite{magnant-crifo:1973}, we had to scale these radiances.  
We estimate (see below) that the radiances would have decreased
by about a factor of 2.4 at 1.1\rsun, and have therefore 
scaled the predicted values by this amount.

We have chosen the scan No. 355 as an example of a quiet region, 
noting that the 1 MK lines such as  \ion{Fe}{x} are quite weak, but 
there are regions where they are even weaker.
The resulting observed and predicted radiances are shown in 
Table~\ref{tab:big_table}, in the third intensity column. 
Again we expected an agreement within a factor of 2, but the
agreement is much better, for the few lines observed and emitted around 1 MK. 
This provides confidence in the present estimates.
It is also interesting to note that the radiance of 
 \ion{S}{xii} is well reproduced, i.e. also the 
scan No. 355 of \cite{magnant-crifo:1973} is consistent with 
photospheric abundances.

\subsection{The QR simulation}

We have chosen the radiances of the scan No. 346 as reported by 
\cite{magnant-crifo:1973},  taken 
close to 1.1\rsun, as representative of a more quiet region.
Indeed, the hot lines such as \ion{Fe}{xv}, \ion{Ni}{xv} are absent
in the spectra. The radiances of the lines formed between 1 and 2 MK
are only slightly  higher than  those in quiet Sun regions, but 
the cooler lines such as  \ion{Fe}{ix} are unusually strong.
Therefore, a simple isothermal assumption was not sufficient.
We have chosen a two-temperature model with 
log $T$ [K]=5.8,6.18 and 
log EM=26.5,26.74,  calculated with a constant density of  1$\times 10^8$ (cm$^{-3}$).
The results of the simulation are shown in the second intensity column
in Table~\ref{tab:big_table}. 
We have adopted in this case  our `coronal' abundances, 
obtained from X-ray and EUV observations of 3 MK emission
in the cores of active regions \citep{delzanna:2013_multithermal,delzanna_mason:2014}.
The  abundances of low-FIP elements are 
 increased by a factor of about 3.2, compared to photospheric.


\subsection{The hot AR simulation}

As we have mentioned, the best AR observation was the 1952 one by 
Lyot. Unfortunately, the height of the slit on the west limb where the active region
was is not certain. The authors have reported initially 1.05\rsun,
then about 1.1\rsun. 
Our understanding of active region structure is much improved form the 
observations from SoHO, Hinode and SDO.
It is  well established that the cores of active regions
are organised into cooler 1 MK loops that are 
 nearly isothermal \cite[see, e.g.][]{delzanna:03}, 
an unresolved 1--2 MK emission \cite[see, e.g.][]{delzanna_mason:03}, and nearly
isothermal 2.5--3 MK core loops \citep[see, e.g.][]{rosner_etal:78}.
Modelling the 1952 AR observations is therefore quite complex,
as along the line of sight there are always different temperature components.
The purpose of this paper is not to provide detailed modelling as was 
e.g. carried out for this specific observation by \cite{mason:1975},
but rather provide an approximate estimate for the lines in the 
far infrared. We have chosen a multi-temperature model 
with log $T$ [K]=6,6.2,6.45,6.6 and 
log EM=26.8,27.2,28.2,27.9, and constant
density of 1$\times 10^9$ (cm$^{-3}$), which we know is typical
for the cores of an active region.
As all the active region cores seem to show a FIP effect,
we have adopted for these simulations our 
coronal values  \citep{delzanna:2013_multithermal,delzanna_mason:2014}.
The observed and predicted radiances are shown in Table~\ref{tab:big_table}.
The agreement is not very good, but typically within a factor of two,
which is acceptable.

We also list in Table~\ref{tab:big_table} the radiances
of the few lines which show a significant difference when the 
calculations are carried out without photoexcitation.
This is to point out 
which lines  are particularly sensitive to photoexcitation.
We note that 
\ion{Ca}{xv} is a particular case: the yellow line is predicted to increase 
due to the photoexcitation by a factor of three.  
The actual increase could be quite different, depending on the 
geometry (recall that we have assumed a simple uniform distribution).

\begin{table*}[!htbp]
\centering
\caption{Line list of visible and infrared lines.
The measured wavelengths (air, in \AA) in first column are, 
otherwise stated, from \cite{curtis_etal:1965}. 
The number in brackets is the uncertainty, in units of the last digit,
whenever available. The letter in brackets (A,B,C) indicate the 
 \cite{aly_report} classification. The Roman numerals indicate the class.
$^{\rm A}$:  as measured by  Aly \cite{lyot_aly:1955};
$^{\rm AEO}$: as  measured by  Aly, Evans and Orrall \citep{aly_report};
$^{\rm KS}$: as  measured by \cite{kernoa_servajean:1965};
$^{\rm J}$: as  measured by \cite{jefferies_etal:1971};
 $^{\rm J02}$: as  measured by \cite{judge_etal:2002}; 
$^{\rm K}$: as  measured by \cite{kurt:1962};
$^{\rm LA}$: as measured by  \cite{lyot_aly:1955};
$^{\rm L}$: as  measured by \cite{lyot:1939};
$^{\rm M}$: as  measured by \cite{munch_etal:1967};
$^{\rm O}$: as  measured by \cite{olsen_etal:1971};
$^{\rm PM}$: as  measured by \cite{petrie_menzel:1942};
$^{\rm WF}$: \cite{wlerick_fehrenbach:1963};
K93: identification suggested by \cite{kastner:1993};
S77: identification listed in the review by \cite{smitt:1977};
MN77:  identification suggested by \cite{mason_nussbaumer:77};
J98: line listed by \cite{judge:1998}.
The second column gives the vacuum wavelengths,
the third the CHIANTI version 8 wavelengths.
I (AR, QR) are the observed  (and predicted in brackets)
radiances of an active region and a more quiet region. 
The second number within the brackets in the I (AR) column indicates,
if present, the radiance as calculated without photoexcitation.
I(QS) indicates predicted radiances for the quiet Sun. 
The estimated radiances are at  1.1\rsun, in photons cm$^{-2}$ s$^{-1}$ arcsec$^{-2}$.
See text for more details.
\label{tab:big_table} 
}
\begin{tabular}{@{}lllllllll@{}}
\tableline
\tableline
 $\lambda_o$(air) & $\lambda_o$(vac)  &  $\lambda_C$  &  I(AR) &  I (QR) &  I (QS) & Ion & Notes & Transition (lower-upper) \\ 
\tableline

 - & -  & 3000.93 &   (1.4)  & (0.3) & (0.2) & \ion{Ni}{xi} & - & 3s$^{2}$ 3p$^{5}$ 3d $^3$F$_{3}$-3s$^{2}$ 3p$^{5}$ 3d $^1$D$_{2}$ \\
-  & -  & 3002.49 &  (2.2)  & (1.0) &  (0.1) & \ion{Fe}{ix} & - & 3s$^{2}$ 3p$^{5}$ 3d $^3$F$_{3}$-3s$^{2}$ 3p$^{5}$ 3d $^3$D$_{2}$ \\

 -  & - & 3012.41 &   (0.6)  & (0.2) & (0.08) & \ion{Fe}{xi} & - & 3s$^{2}$ 3p$^{3}$ 3d $^3$F$_{3}$-3s$^{2}$ 3p$^{3}$ 3d $^1$G$_{4}$ \\

3021.3$^{\rm J}$ & 3022.2 & 3020.97 & (17)  & (4) & (2) &  \ion{Fe}{x} &  S77 & 3s$^{2}$ 3p$^{4}$ 3d $^4$F$_{9/2}$-3s$^{2}$ 3p$^{4}$ 3d $^2$G$_{9/2}$ \\
   & & & & & & &   (not Fe XII)  & \\

3072.0  & 3072.9 & ?3072.06 &   (13)  & (3.2) & (0.9) &  \ion{Fe}{xii} & S77 & 3s$^{2}$ 3p$^{3}$ $^2$D$_{3/2}$-3s$^{2}$ 3p$^{3}$ $^2$P$_{1/2}$ \\

3124.0 & 3124.9 & 3124.06 &   (7.6)  & (3.7) & (0.4) &  \ion{Fe}{ix} & New & 3s$^{2}$ 3p$^{5}$ 3d $^3$F$_{2}$-3s$^{2}$ 3p$^{5}$ 3d $^1$F$_{3}$ \\

3167.0 & 3167.9  & 3167.57 &   (0.9)  & (0.2) & (0.08) &  \ion{Ni}{xii} & S77 & 3s$^{2}$ 3p$^{4}$ 3d $^4$D$_{7/2}$-3s$^{2}$ 3p$^{4}$ 3d $^4$F$_{9/2}$ \\
   & & & & & & &  (not Cr XI)   & \\

3302.8 & 3303.8 & 3302.37 &   (0.4)  & (0.03) & (0.01) &  \ion{Cr}{ix} & S77: Ni XI & 3s$^{2}$ 3p$^{4}$ $^3$P$_{2}$-3s$^{2}$ 3p$^{4}$ $^1$D$_{2}$ \\

3327.5 (A) & 3328.5 & 3328.46 &  66 (54)  & (1.4) & (0.2) & \ion{Ca}{xii} & - & 2s$^{2}$ 2p$^{5}$ $^2$P$_{3/2}$-2s$^{2}$ 2p$^{5}$ $^2$P$_{1/2}$ \\

3338.5 (A) & 3339.5 & 3339.46 &   (1.5)  & (0.5) & (0.2) &   \ion{Ni}{xi} & New & 3s$^{2}$ 3p$^{5}$ 3d $^3$F$_{3}$-3s$^{2}$ 3p$^{5}$ 3d $^3$D$_{3}$ \\

3355.1 & 3356.1 & 3355.44 &   (1.)  & (0.7) & (0.07) &  \ion{Fe}{ix} & S77 & 3s$^{2}$ 3p$^{5}$ 3d $^3$F$_{4}$-3s$^{2}$ 3p$^{5}$ 3d $^3$D$_{3}$ \\

3388.10$^{\rm L}$(7) (A) & 3389.07 & 3388.9  &  220 (160)  & 27 (24) & 9 (5) & \ion{Fe}{xiii} & - & 3s$^{2}$ 3p$^{2}$ $^3$P$_{2}$-3s$^{2}$ 3p$^{2}$ $^1$D$_{2}$ \\

  &  & 3446.2561  & (2.4) &   -  & - & \ion{K}{xv} & - & 2s$^{2}$ 2p $^2$P$_{1/2}$-2s$^{2}$ 2p $^2$P$_{3/2}$ \\

3454.2 (A) & 3455.2 & 3454.95 &  24 (23)  & 29 (11) & 7 (4) &  \ion{Fe}{x} & MN77 & 3s$^{2}$ 3p$^{4}$ 3d $^4$D$_{7/2}$-3s$^{2}$ 3p$^{4}$ 3d $^4$F$_{9/2}$ \\

3471.6 (B) & 3472.6 & 3472.49 &   (1.9)  & 5 (0.9) & (0.09) &  \ion{Fe}{ix} & S77 & 3s$^{2}$ 3p$^{5}$ 3d $^3$F$_{2}$-3s$^{2}$ 3p$^{5}$ 3d $^3$D$_{2}$ \\

3488.5  &  3489.5  & ?3484.80 & (0.5)  & (0.2) & (0.07) & \ion{Fe}{xi} & New & 3s$^{2}$ 3p$^{3}$ 3d $^3$F$_{4}$-3s$^{2}$ 3p$^{3}$ 3d $^1$G$_{4}$ \\

3488.5 & 3489.5 & ?3489.92 &   (0.1)  & (1.7) & - &  ? \ion{Mg}{vi} & New & 2s$^{2}$ 2p$^{3}$ $^2$D$_{3/2}$-2s$^{2}$ 2p$^{3}$ $^2$P$_{3/2}$ \\

        &      &    ?3489.9  & ()   & (2.7)  & - &   ?\ion{Mg}{vi} & New & 2s$^{2}$ 2p$^{3}$ $^2$D$_{3/2}$-2s$^{2}$ 2p$^{3}$ $^2$P$_{3/2}$ \\

3502.5 & 3503.5 & 3503.25  &   (8$\times$10$^{-2}$)  & (2) & - &   ?\ion{Mg}{vi} & S77: Cu XIII & 2s$^{2}$ 2p$^{3}$ $^2$D$_{3/2}$-2s$^{2}$ 2p$^{3}$ $^2$P$_{1/2}$ \\

3533.6 (A) & 3534.6  & 3533.82 &  7 (5)  & 4 (1.3) & (0.6) &  \ion{Fe}{x} & MN77 & 3s$^{2}$ 3p$^{4}$ 3d $^4$F$_{7/2}$-3s$^{2}$ 3p$^{4}$ 3d $^2$G$_{7/2}$ \\
   & & & & & & &   (not V X)  & \\

3577.1 (I,II) (B) & 3578.1 & ?3575.4  &  (1.6)  & 3.4 (0.4) & (0.2) &  \ion{Fe}{x} & S77 & 3s$^{2}$ 3p$^{4}$ 3d $^4$F$_{7/2}$-3s$^{2}$ 3p$^{4}$ 3d $^2$G$_{9/2}$ \\

3601.1 (A) & 3602.1 & 3602.2539 &  77 (120, 110)  & -  & - & \ion{Ni}{xvi} & - & 3s$^{2}$ 3p $^2$P$_{1/2}$-3s$^{2}$ 3p $^2$P$_{3/2}$ \\

3642.7 (A) & 3643.7 & 3644.1 &  10 (11.)  & 6 (5.6) & 1.3 (0.6) &   \ion{Fe}{ix} & S77 & 3s$^{2}$ 3p$^{5}$ 3d $^3$F$_{3}$-3s$^{2}$ 3p$^{5}$ 3d $^1$D$_{2}$ \\
   & & & & & & &   (not Ni XIII)  & \\

3645.9$^{\rm AEO}$ (D)  &   & 3646.85  & (0.26) &   -  & - & ? \ion{Ca}{xvii} & New & 2s 2p $^3$P$_{1}$-2s 2p $^3$P$_{2}$ \\

3800.8 (I) (A) & 3801.9 & 3802.1 &  14 (13)  & 10 (9)  & 6 (0.9) &  \ion{Fe}{ix} & S77 & 3s$^{2}$ 3p$^{5}$ 3d $^3$F$_{3}$-3s$^{2}$ 3p$^{5}$ 3d $^3$D$_{3}$ \\
   & & & & & & &   (not Co XII)  & \\

3986.8 (A) & 3987.9 & 3988.0 &  21 (21)  & 15 (7.5) & 8 (3) & \ion{Fe}{xi} & - & 3s$^{2}$ 3p$^{4}$ $^3$P$_{1}$-3s$^{2}$ 3p$^{4}$ $^1$D$_{2}$ \\

4087.1 (A) &  4088.3 & 4087.5  &  105 (72, 49)  & (0.3) & (0.02) & \ion{Ca}{xiii} & - & 2s$^{2}$ 2p$^{4}$ $^3$P$_{2}$-2s$^{2}$ 2p$^{4}$ $^3$P$_{1}$ \\
\\

? 4111$^{\rm KS}$ (III?)  &  -  & 4114.0 &   (1.2)  & (0.3) & (0.14) & ? \ion{Fe}{x} & New & 3s$^{2}$ 3p$^{4}$ 3d $^4$F$_{5/2}$-3s$^{2}$ 3p$^{4}$ 3d $^2$G$_{7/2}$ \\

4231.2 (A) & 4232.4 & 4232.1 &  68 (20,18)  & 7 (9) & (2.4) &  \ion{Ni}{xii} & - & 3s$^{2}$ 3p$^{5}$ $^2$P$_{3/2}$-3s$^{2}$ 3p$^{5}$ $^2$P$_{1/2}$ \\

? 4251$^{\rm AEO}$ (C)  & -  &  4249.9 &   (1,0.7)  & (0.2) & (0.04) & ? \ion{K}{xi} & New & 2s$^{2}$ 2p$^{5}$ $^2$P$_{3/2}$-2s$^{2}$ 2p$^{5}$ $^2$P$_{1/2}$ \\

4311.8 (C) & 4313.0 & 4312.8 &   (4.5)  & 5 (1.3) & (0.5) &   \ion{Fe}{x} &  MN77 & 3s$^{2}$ 3p$^{4}$ 3d $^4$F$_{9/2}$-3s$^{2}$ 3p$^{4}$ 3d $^2$F$_{7/2}$ \\
   & & & & & & &  (not V X)   & \\

4359.4 (A) & 4360.6 & 4360.4 &   (8.5)  & 9 (4.2) & (0.4) & \ion{Fe}{ix} &  S77 & 3s$^{2}$ 3p$^{5}$ 3d $^3$F$_{2}$-3s$^{2}$ 3p$^{5}$ 3d $^1$D$_{2}$ \\

 4413 (A) &   & 4413.8 &  83 (42, 26)  & 6 (-) & - & \ion{Ar}{xiv} & - & 2s$^{2}$ 2p $^2$P$_{1/2}$-2s$^{2}$ 2p $^2$P$_{3/2}$ \\

4566.2 (A) & 4567.5  & 4567.5  &   (6)  & (4.3) & (1.4) &  \ion{Fe}{xi} & MN77 & 3s$^{2}$ 3p$^{3}$ 3d $^3$F$_{4}$-3s$^{2}$ 3p$^{3}$ 3d $^3$G$_{5}$ \\

   & & & & & & &  (not Cr IX)   & \\

4585.3 (D) &  4586.6 & ?4588.6 &   (4)  & (2.8) & (0.3) &  \ion{Fe}{ix} & S77 & 3s$^{2}$ 3p$^{5}$ 3d $^3$F$_{2}$-3s$^{2}$ 3p$^{5}$ 3d $^3$D$_{3}$ \\

4744$^{\rm A}$ (D) &  4745.3 & ?4750.1 &   (7.5)  & -  & - &  ?\ion{Ni}{xvii} & New & 3s 3p $^3$P$_{1}$-3s 3p $^3$P$_{2}$ \\

  &  & 5079. &   (1.8)  &  (1) & (0.3) & \ion{Fe}{xi} & - & 3s$^{2}$ 3p$^{3}$ 3d $^3$F$_{4}$-3s$^{2}$ 3p$^{3}$ 3d $^3$G$_{4}$ \\


5116.03$^{\rm L}$(2) (A) & 5117.5 & 5117.2 & 114  (20, 17)  & (3.7) & (0.6) & \ion{Ni}{xiii} & J98:1 & 3s$^{2}$ 3p$^{4}$ $^3$P$_{2}$-3s$^{2}$ 3p$^{4}$ $^3$P$_{1}$ \\

  &   & 5278.43 &   (5, 2)  & (0.2) & - & \ion{K}{xii} & - & 2s$^{2}$ 2p$^{4}$ $^3$P$_{2}$-2s$^{2}$ 2p$^{4}$ $^3$P$_{1}$ \\

5302.86$^{\rm L}$ (2) (A) &  5304.34 & 5304.5 &  1481 (1400, 1300)  & 144 (140) & 16 (11) & \ion{Fe}{xiv} &  J98:144 & 3s$^{2}$ 3p $^2$P$_{1/2}$-3s$^{2}$ 3p $^2$P$_{3/2}$ \\

5446.0$^{\rm LA}$ (A) & 5447.5 & 5446.0 &  95 (33,30)  & -  & - & \ion{Ca}{xv} & J98:0.2 & 2s$^{2}$ 2p$^{2}$ $^3$P$_{1}$-2s$^{2}$ 2p$^{2}$ $^3$P$_{2}$ \\


? 5533.4 (I) &  5534.9  & 5535.6  &     (7.5,4)  & (8) & (5) &  S77: \ion{Ar}{x} & (J98:2) & 2s$^{2}$ 2p$^{5}$ $^2$P$_{3/2}$-2s$^{2}$ 2p$^{5}$ $^2$P$_{1/2}$ \\

? 5537$^{\rm AEO}$ (I) (A) & 5538  &    5535.6  &  32 (7.5,4)  & (8) & (5) &  \ion{Ar}{x} & (J98:2) & 2s$^{2}$ 2p$^{5}$ $^2$P$_{3/2}$-2s$^{2}$ 2p$^{5}$ $^2$P$_{1/2}$ \\

? 5539.1 (I) (A)  & 5540.6  &  5538.9 &  -  (2)  & - (0.5) & (0.2) & S77: \ion{Fe}{x} &  & 3s$^{2}$ 3p$^{4}$ 3d $^4$F$_{7/2}$-3s$^{2}$ 3p$^{4}$ 3d $^2$F$_{7/2}$ \\
 & & &  & & & (not \ion{Ar}{x}) &  &  \\


5694.42$^{\rm L}$ (5) (A) & 5696.0  & 5696.4 &  186 (190, 65)  &  -  & - & \ion{Ca}{xv} & J98:3.4 & 2s$^{2}$ 2p$^{2}$ $^3$P$_{0}$-2s$^{2}$ 2p$^{2}$ $^3$P$_{1}$ \\

 -  &  -  & 5743.8  &   (9, 6)  & - & - & \ion{Cl}{xiii} & - & 2s$^{2}$ 2p $^2$P$_{1/2}$-2s$^{2}$ 2p $^2$P$_{3/2}$ \\

5944$^{\rm AEO}$  (B)  &   & 5945.3  & (0.8) &   -  &  & ? \ion{Ar}{xv} & - & 2s 2p $^3$P$_{1}$-2s 2p $^3$P$_{2}$ \\

\tableline
\end{tabular}
\end{table*}

\addtocounter{table}{-1}
\begin{table*}
\centering
\caption{Contd. 
\label{tab:obs}}
\begin{tabular}{@{}lllllllll@{}}
 \tableline
 \tableline
 $\lambda_o$(air) & $\lambda_o$(vac)  &  $\lambda_C$  &  I(AR)  &  I (QR) &  I (QS) & Ion & Notes & Transition (lower-upper) \\ 
 \tableline

6374.56$^{\rm WF}$ (A) & 6376.32 & 6376.3 & 163 (310, 280)  & 211 (200) & 73 (57) & \ion{Fe}{x} &  J98:99 & 3s$^{2}$ 3p$^{5}$ $^2$P$_{3/2}$-3s$^{2}$ 3p$^{5}$ $^2$P$_{1/2}$ \\

? 6622.99$^{\rm WF}$ &  6624.82 & 6624.3 &   (1.7)  & (0.7) & (0.3) & ? \ion{Fe}{xi} & - & 3s$^{2}$ 3p$^{3}$ 3d $^3$D$_{3}$-3s$^{2}$ 3p$^{3}$ 3d $^3$F$_{4}$ \\

 &  & ? 6672.5  & (2.4, 1.9) &   - & - & \ion{K}{xiv} & - & 2s$^{2}$ 2p$^{2}$ $^3$P$_{1}$-2s$^{2}$ 2p$^{2}$ $^3$P$_{2}$ \\

6701.47$^{\rm WF}$ (A) & 6703.32 & 6703.5 & 216 (140, 99) & (0.5) & (0.03) &  \ion{Ni}{xv} & S77 & 3s$^{2}$ 3p$^{2}$ $^3$P$_{0}$-3s$^{2}$ 3p$^{2}$ $^3$P$_{1}$ \\

6917.0$^{\rm WF}$ & 6918.91 & 6918.0  &  (18, 8)  & 4 (6.6) & (3) &  \ion{Ar}{xi} & S77. J98:3.2 & 2s$^{2}$ 2p$^{4}$ $^3$P$_{2}$-2s$^{2}$ 2p$^{4}$ $^3$P$_{1}$ \\

7059.59$^{\rm WF}$ & 7061.54  & 7062.1 &   (200)  & (1.1) & (0.08) & \ion{Fe}{xv} & J98:3.5 & 3s 3p $^3$P$_{1}$-3s 3p $^3$P$_{2}$ \\

 &  & 7548.3 &   (13, 4)  &   -  & - & \ion{K}{xiv} & - & 2s$^{2}$ 2p$^{2}$ $^3$P$_{0}$-2s$^{2}$ 2p$^{2}$ $^3$P$_{1}$ \\

7611.0 (II) & 7613.1 & 7613.1 &   (260, 180)  & 23 (14) & 3.4 (2.7) &  \ion{S}{xii} & S77 & 2s$^{2}$ 2p $^2$P$_{1/2}$-2s$^{2}$ 2p $^2$P$_{3/2}$ \\

7891.89$^{\rm WF}$ & 7894.06  & 7894.0 & (340,290)  & 177 (420) & 66 (93) & \ion{Fe}{xi} & J98:85 & 3s$^{2}$ 3p$^{4}$ $^3$P$_{2}$-3s$^{2}$ 3p$^{4}$ $^3$P$_{1}$ \\

8024.2$^{\rm L}$(1) & 8026.3 & 8026.3  &   (57)  & (6$\times$10$^{-2}$) & - &  \ion{Ni}{xv} & - & 3s$^{2}$ 3p$^{2}$ $^3$P$_{1}$-3s$^{2}$ 3p$^{2}$ $^3$P$_{2}$ \\

  &   & 8339.6 &   (20)  &  -  & - & \ion{Ar}{xiii} & J98:0.5 & 2s$^{2}$ 2p$^{2}$ $^3$P$_{1}$-2s$^{2}$ 2p$^{2}$ $^3$P$_{2}$ \\

  &  & 9219. &   (0.5)  & (0.1) & - & \ion{Cl}{x} & - & 2s$^{2}$ 2p$^{4}$ $^3$P$_{2}$-2s$^{2}$ 2p$^{4}$ $^3$P$_{1}$ \\

 &  & 9917 &   (8.5)  & (4) & (1.5) & \ion{S}{viii} & J98:1.2 & 2s$^{2}$ 2p$^{5}$ $^2$P$_{3/2}$-2s$^{2}$ 2p$^{5}$ $^2$P$_{1/2}$ \\

 &  & 9919 &   (4.3)  & (0.5) & (0.1) & \ion{Fe}{xiii} & - & 3s$^{2}$ 3p 3d $^3$F$_{3}$-3s$^{2}$ 3p 3d $^3$F$_{4}$ \\

  &  & 9978 &    (1.1)  & (0.4) & (0.2) & \ion{Mn}{x} & - & 3s$^{2}$ 3p$^{4}$ $^3$P$_{2}$-3s$^{2}$ 3p$^{4}$ $^3$P$_{1}$ \\

 &  & 10143 &   (38)  &   -  & - & \ion{Ar}{xiii} & J98:8 & 2s$^{2}$ 2p$^{2}$ $^3$P$_{0}$-2s$^{2}$ 2p$^{2}$ $^3$P$_{1}$ \\

 &  & 10301 &   (13)  &  -  & - & \ion{S}{xiii} & - & 2s 2p $^3$P$_{1}$-2s 2p $^3$P$_{2}$ \\

 &  & 10651 &   (2)  &  -  & - & \ion{Cl}{xii} & - & 2s$^{2}$ 2p$^{2}$ $^3$P$_{1}$-2s$^{2}$ 2p$^{2}$ $^3$P$_{2}$ \\


10746.80$^{\rm L}$(15) & 10749.7 & 10749. &   (4.3$\times$10$^{2}$)  & (240) & (35) & \ion{Fe}{xiii} & J98:285 & 3s$^{2}$ 3p$^{2}$ $^3$P$_{0}$-3s$^{2}$ 3p$^{2}$ $^3$P$_{1}$ \\
10797.95$^{\rm L}$ (15) & 10801. & 10801. &   (3.4$\times$10$^{2}$)  & (140) & (21) & \ion{Fe}{xiii} & J98:67 & 3s$^{2}$ 3p$^{2}$ $^3$P$_{1}$-3s$^{2}$ 3p$^{2}$ $^3$P$_{2}$ \\

   & 12660 & ? 12524 &  224$^{\rm O}$ (14)  &  (4.7) & (4.7)  &  \ion{S}{ix} & J98:13 & 2s$^{2}$ 2p$^{4}$ $^3$P$_{2}$-2s$^{2}$ 2p$^{4}$ $^3$P$_{1}$ \\
   & & & & & & & (not Fe XIV) & \\

   &  & 12793 &   (2.7)  &   -  & - & \ion{Ni}{xiv} & - & 3s$^{2}$ 3p$^{3}$ $^2$D$_{3/2}$-3s$^{2}$ 3p$^{3}$ $^2$D$_{5/2}$ \\

  &  & 13837 &   (4)  &  -  & - & \ion{Cl}{xii} & - & 2s$^{2}$ 2p$^{2}$ $^3$P$_{0}$-2s$^{2}$ 2p$^{2}$ $^3$P$_{1}$ \\

  &   & 13928 &   (50)  & (7) & (3.5) & \ion{S}{xi} & J98:6 & 2s$^{2}$ 2p$^{2}$ $^3$P$_{1}$-2s$^{2}$ 2p$^{2}$ $^3$P$_{2}$ \\

   & 14305$^{\rm M}$ & 14305 &   262$^{\rm O}$ (170)  & 51$^{\rm M}$ (130) & (37) & \ion{Si}{x} & J98:21 & 2s$^{2}$ 2p $^2$P$_{1/2}$-2s$^{2}$ 2p $^2$P$_{3/2}$ \\

 & 19220$^{\rm O}$ & 19201.2 & $<$159$^{\rm O}$  (45)  &  (12) & (5.6) & \ion{S}{xi} & K93; J98:21  & 2s$^{2}$ 2p$^{2}$ $^3$P$_{0}$-2s$^{2}$ 2p$^{2}$ $^3$P$_{1}$ \\

   & & & & & & &   (not Si XI)  & \\

 &  & 19350 &   (16)  & (5) & (1) & \ion{Si}{xi} & J98:0.9 & 2s 2p $^3$P$_{1}$-2s 2p $^3$P$_{2}$ \\

 &  & 19482 &   (3.3)  & (1) & (0.5) & \ion{Fe}{x} & - & 3s$^{2}$ 3p$^{4}$ 3d $^4$F$_{9/2}$-3s$^{2}$ 3p$^{4}$ 3d $^4$F$_{7/2}$ \\

 &  & 19630.0 &   ()  & (18) & (0.01) & \ion{Si}{vi} & - & 2s$^{2}$ 2p$^{5}$ $^2$P$_{3/2}$-2s$^{2}$ 2p$^{5}$ $^2$P$_{1/2}$ \\

 &  & 20450 &   (3.7)  & (3.3) & (1) & \ion{Al}{ix} & - & 2s$^{2}$ 2p $^2$P$_{1/2}$-2s$^{2}$ 2p $^2$P$_{3/2}$ \\

 &  & 22063 &   (16)  & (18) & (3.6) &  \ion{Fe}{xii} & - & 3s$^{2}$ 3p$^{3}$ $^2$D$_{3/2}$-3s$^{2}$ 3p$^{3}$ $^2$D$_{5/2}$ \\

 &  & 22183 &   (7)  & (11) & (0.9) & \ion{Fe}{ix} & J98:31 & 3s$^{2}$ 3p$^{5}$ 3d $^3$F$_{3}$-3s$^{2}$ 3p$^{5}$ 3d $^3$F$_{2}$ \\

 &  & 22650 &   (9)  &  -  & - & \ion{Ca}{xiii} & J98:0.08 & 2s$^{2}$ 2p$^{4}$ $^3$P$_{1}$-2s$^{2}$ 2p$^{4}$ $^3$P$_{0}$ \\

 &  & 24307 &   (1)  & (1.5) & (0.5) & \ion{Ni}{xi} & - & 3s$^{2}$ 3p$^{5}$ 3d $^3$F$_{4}$-3s$^{2}$ 3p$^{5}$ 3d $^3$F$_{3}$ \\

 &  & 24826 &   (5)  & (41) & (0.3) & \ion{Si}{vii} & J98:6 & 2s$^{2}$ 2p$^{4}$ $^3$P$_{2}$-2s$^{2}$ 2p$^{4}$ $^3$P$_{1}$ \\

 &  & 25846 &   (34)  & (41) & (12) & \ion{Si}{ix} & J98:27 & 2s$^{2}$ 2p$^{2}$ $^3$P$_{1}$-2s$^{2}$ 2p$^{2}$ $^3$P$_{2}$ \\

   & 27470$^{\rm O}$ & ?27541  &  $<$325$^{\rm O}$ (0.6)  &  (0.3) & (0.1) & ? \ion{Al}{x} & - & 2s 2p $^3$P$_{1}$-2s 2p $^3$P$_{2}$ \\

   &    & ? 27159 & (0.14) &   (0.2) & 0.15  &  \ion{P}{x} & - & 2s$^{2}$ 2p$^{2}$ $^3$P$_{0}$-2s$^{2}$ 2p$^{2}$ $^3$P$_{1}$ \\

  &  & 28563. &   (5)  & (17) & (1.1) & \ion{Fe}{ix} & J98:9 & 3s$^{2}$ 3p$^{5}$ 3d $^3$F$_{4}$-3s$^{2}$ 3p$^{5}$ 3d $^3$F$_{3}$ \\

   & 30275$^{\rm M}$ & ?30284.7 & 25$^{\rm M}$  $<$357$^{\rm O}$ (9) & (34.) & (0.7) & \ion{Mg}{viii} & J98:16  & 2s$^{2}$ 2p $^2$P$_{1/2}$-2s$^{2}$ 2p $^2$P$_{3/2}$ \\


   &  & 37551. &   (1)  & (0.9) & (0.7) & \ion{S}{ix} & J98:0.4 & 2s$^{2}$ 2p$^{4}$ $^3$P$_{1}$-2s$^{2}$ 2p$^{4}$ $^3$P$_{0}$ \\

39343.4$^{\rm J02}$   &  & 39277. &   (10)  & (29) & (6) & \ion{Si}{ix} & J98:31 & 2s$^{2}$ 2p$^{2}$ $^3$P$_{0}$-2s$^{2}$ 2p$^{2}$ $^3$P$_{1}$ \\

  &  & 54466  &   ()  & (18.) & (0.07) & \ion{Fe}{viii} & - & 3s$^{2}$ 3p$^{6}$ 3d $^2$D$_{3/2}$-3s$^{2}$ 3p$^{6}$ 3d $^2$D$_{5/2}$ \\
  &  & 55033  &   ()  & (16.) & (0.01) & \ion{Mg}{vii} & J98: 1.8 & 2s$^{2}$ 2p$^{2}$ $^3$P$_{1}$-2s$^{2}$ 2p$^{2}$ $^3$P$_{2}$ \\


  &  & 64935  &   ()  & (3.8) & (0.02) & \ion{Si}{vii} & J98:0.3 & 2s$^{2}$ 2p$^{4}$ $^3$P$_{1}$-2s$^{2}$ 2p$^{4}$ $^3$P$_{0}$ \\


  &  & 90090  &   ()  & (3.3) & - & \ion{Mg}{vii} & J98:0.5 & 2s$^{2}$ 2p$^{2}$ $^3$P$_{0}$-2s$^{2}$ 2p$^{2}$ $^3$P$_{1}$ \\

\tableline
\end{tabular}
\end{table*}

\begin{table}[!htbp]
\centering
\caption{List of the observed lines with no atomic data available.
The few suggested  identifications are noted, as
well as the \cite{aly_report} classification: (A,B,C).
See Table~2 for the legend.
\label{tab:obs_not}
}
\begin{tabular}{@{}llllllllll@{}}
\tableline
\tableline
 $\lambda_o$(air) & $\lambda_o$(vac)  &  Notes   \\ 
\tableline

3328.6$^{\rm PM}$  &  &    \\
 3466.9$^{\rm AEO}$ &  &  (C)   \\
3573.2$^{\rm AEO}$ &  &  (B) \\
3685.5  & 3686.6  &     Mn XIII (C) \\

 3781.4$^{\rm AEO}$ &  &   (C) \\

 3980.9$^{\rm PM}$  &  &    \\

 3925$^{\rm AEO}$ &  &    (C)  \\
 3946$^{\rm AEO}$ &  &    (C)  \\
 3959.4$^{\rm AEO}$ &  &   (C)   \\
 3972.3$^{\rm AEO}$ &  &   (C)   \\

3996.8 (I), 3998$^{\rm LA}$  & 3997.9 &  S77: Cr XI (B)  \\

 4003.5$^{\rm PM}$  &  &    \\

 4056.3$^{\rm PM}$  &  &    \\

 4085.6$^{\rm PM}$ &  &    \\

 4096$^{\rm AEO}$ &  &   (C)  \\

 4140$^{\rm A}$  &  & (C) \\

 4170.8$^{\rm PM}$,4170$^{\rm A}$  &  & (C) \\
 4214$^{\rm AEO}$ &  &   (C)   \\

 4272.9$^{\rm PM}$  &  &    \\

 4316.2$^{\rm AEO}$ &  &  (C)    \\
 4322.7$^{\rm AEO}$ &  &   (C)   \\

4351$^{\rm KS}$ (III), 4348$^{\rm AEO}$  &  & ? Co XV  (C) \\

4418$^{\rm A, DP}$  &  &  (C) \\

 4568.7$^{\rm AEO}$ &  &   (C)   \\

 4578.2$^{\rm AEO}$ &  &   (C)   \\

 4668.9$^{\rm AEO}$ &  &   (B)   \\

 4953$^{\rm AEO}$ &  &   (C)   \\

 5105.4$^{\rm AEO}$ &  &   (C)   \\

5130.7$^{\rm AEO}$ &  &   (B)   \\

 5483$^{\rm AEO}$ &  &   (C)  \\
 5499.2$^{\rm AEO}$ &  &  (B)    \\

? 5537$^{\rm PM, AEO}$  &  & ? \ion{Ar}{x} (A)  \\

 5560$^{\rm AEO}$ &  &   (C)  \\ 
 5618.3$^{\rm AEO}$ &  &  (C)  \\ 

 5907$^{\rm AEO}$ &  &  (B)   \\

 5899.1$^{\rm PM}$  &    & \\

 5912.4$^{\rm PM}$  &  &  \\
 5926$^{\rm AEO}$ &  &  (B)  \\

 5937.1$^{\rm PM}$  &  &  \\

 6294.9$^{\rm PM}$  &  &    \\

 6272.06$^{\rm WF}$  & 6273.80 &  \\

6304.60$^{\rm WF}$  & 6306.34  &  ?Co XII  \\

6336.9$^{\rm PM}$  &  &   \\

 6388$^{\rm AEO}$ &  &   (B)    \\

 6404$^{\rm AEO}$ &  &   (B)   \\

 6436.0$^{\rm WF}$ & 6437.78  &    \\

 6456$^{\rm AEO}$ &  &   (B)   \\

 6476$^{\rm AEO}$ &  &    (B)  \\

6513.0$^{\rm PM}$, 6511$^{\rm AEO}$ &  &   (B)   \\

 6524.1$^{\rm PM}$  &  &    \\

6535.96$^{\rm WF}$(I,II), 6535.4$^{\rm A}$ & 6537.77  & ? Mn XIII (B)  \\

6601.93$^{\rm WF}$ &  6603.75 &     \\

 7092.0$^{\rm WF}$ &  7093.96 &    \\

 \tableline
\end{tabular}
 \end{table}

\addtocounter{table}{-1}
\begin{table}[!htbp]
\centering
\caption{Contd. }
\begin{tabular}{@{}llllllllll@{}}
\tableline
\tableline
 $\lambda_o$(air) & $\lambda_o$(vac)  &  Notes   \\ 
\tableline

7143.90$^{\rm WF}$ &  7145.87 &   \\
7160.75$^{\rm WF}$ &  7162.72 &    \\
7212.76$^{\rm WF}$ & 7214.75  &    \\

7852.9$^{\rm WF}$ & 7855.06 &    \\

7956$^{\rm K}$&   &   \\

 8077.0$^{\rm WF}$ & 8079.22 &   \\

8153.8 (I) & 8156.0   &   ?(S77: Cr XII) \\

 8425.08$^{\rm WF}$ & 8427.40 &    \\
 8427.34$^{\rm WF}$ & 8429.66 &    \\
 8429.10$^{\rm WF}$ & 8431.42 &    \\
 8475.66$^{\rm WF}$ & 8477.99  &    \\
 8477.08$^{\rm WF}$ & 8479.41 &    \\
 8493.09$^{\rm WF}$ & 8495.42 &    \\

11304$^{\rm K}$&   &    \\
11355$^{\rm K}$&   &    \\
11386$^{\rm K}$&   &    \\
11585$^{\rm K}$&   &    \\

 & 15230$^{\rm O}$  & (K93: Cr XI ?)   \\

 & 15280$^{\rm O}$  & (K93: Ti VI ?)   \\

 & 17210$^{\rm O}$  &  \\

 &  18560$^{\rm O}$  & ?Cr XI  \\

 \tableline

\end{tabular}
 \end{table}

\begin{table}[!htbp]
\centering
\caption{Some of the main density diagnostic ratios for the 
moderately quiet solar corona. The ion, the approximate wavelengths
and temperature of formation $T$ of the lines is shown.
In the last column, we show the range of densities where the 
ratios are useful.
\label{tab:ne_table} 
}
\begin{tabular}{@{}lllllllll@{}}
\tableline
\tableline
 Ion &  $\lambda_1$ & $\lambda_2$ & $T$ & log $N_{\rm e}$ \\
     &     \AA &  \AA & MK  &      cm$^{-3}$ \\ 
\tableline

\ion{Fe}{ix} &   3124 &   3355  &   1  & 8.5--11 \\ 
\ion{Fe}{ix} &    3644 & 3802   &   1  & 8--11 \\ 
\ion{Fe}{ix} &    4360 & 4588 &   1  & 8--11 \\ 
\ion{Fe}{ix} &    22183 & 28563 &  1 &  7--9.5 \\ 

\ion{Fe}{x} &  6376 & 3454  &  1 &  9--11 \\
\ion{Fe}{x} &  3021. & 3454  &  1 &  9--11 \\

\ion{Fe}{xi} &   3988. & 7894 &  1.5 &  9--11 \\

 \ion{Si}{ix} & 25846 & 39277 &   1 & 8--9 \\
\ion{S}{xi} & 13928 & 19220  &   2  & 8--10 \\ 

\ion{Fe}{xiii} & 10801 & 10749 & 2 & 7--11  \\

\ion{Ar}{xiii} & 8339 &   10143  &   3  & 9.5--11.5 \\ 

 \ion{Ca}{xv} &  5446 &  5696 & 4.5 & 8--10.5 \\
\tableline
\end{tabular}
\end{table}

\subsection{A few remarks about the Tables and the results}

First, we summarise the contents of Table~\ref{tab:big_table}.
The first column gives the observed wavelength in air.
The uncertainties and Lyot's class of the line is given, whenever provided.
We also list the intensity class  (A,B,C,D)
as reported by \cite{aly_report}. We recall that the lines with class A or B are
certain, C almost certain and D dubious.
A question mark is added when either a line is dubious or the 
wavelength is probably inaccurate.

The second column gives the vacuum wavelengths.
The third column gives the vacuum CHIANTI wavelengths. A question mark
is added to indicate significant disagreement with the values in second column.
The fourth column gives the few observed radiances in an active region (AR), and 
in brackets the predicted ones, with and without photoexcitation
included.
The fifth and sixth columns give observed and predicted radiances for 
a more quiet corona (QR) and the quiet Sun (QS).

The next column indicates the ion. A question mark is added to indicate
a tentative assignment, sometimes previously suggested, which we have been
unable to confirm with the present data.
The following column indicates previous identifications
whenever  a line in the original list was incorrectly identified.

It also displays if the line was in the list of potentially
interesting ones in \cite{judge:1998}, together with its predicted radiance. 
Very good agreement with our estimated radiances of the quiet Sun case (QS) 
can be seen in several cases. This is reassuring, considering that 
\cite{judge:1998}  obtained radiances using a different method,
i.e. an integral along the line of sight (at a distance of 1.1\rsun).
However,  the radiances of the hotter and cooler
lines are somewhat higher in  \cite{judge:1998} simulation. This can be explained by the fact the 
\cite{judge:1998} assumed a constant distribution of plasma from 
 log$_{10}$ $T$ [K]=5.9 to 6.5, at all heights, unlike our 
isothermal  approximation at log $T$[K]=6.14, i.e. 1.38 MK.
In some instances, a few ratios as reported by \cite{judge:1998} 
are incompatible with the present estimates though.
For example, the ratio of the two  \ion{Fe}{xiii}  10750, 10801~\AA\
lines is consistent with a lower density (about 4 $\times$ 10$^7$ cm$^{-3}$)
than our assumed value. The same holds for the  \ion{S}{xi} 13928 vs. the 19220~\AA\ line
(see Figure~\ref{fig:densities} for the CHIANTI v.8 theoretical ratio).

We have chosen not to clog the Table with multiple entries.
Among the observed wavelengths, we list what we think is the best observed wavelength
from the assessed literature. Details on the various measurements 
can be found in the references.

In general, we have listed the identifications as provided by 
\cite{curtis_etal:1965} for the visible lines and as 
provided by \cite{olsen_etal:1971} for the infrared lines, and noted in the Table
when incorrect identifications were listed. 
As previously mentioned, there is a large number of publications
where the original identifications of the various lines have been 
proposed. 
Within the Table, we simply note if a line was previously 
 identified by \cite{mason_nussbaumer:77} or by others, as listed in 
\cite{smitt:1977}, where a nice summary of the identifications known 
at the time was provided.

More information on the original identifications 
and an assessment of the experimental
energies for each ion are provided in the original 
publications which are referenced in the CHIANTI database.

 Table~\ref{tab:big_table} lists only the lines for which 
we currently have atomic data for in CHIANTI version 8.
It includes both observed lines and a few that have not been observed
but are potentially observable. 

On the other hand, Table~\ref{tab:obs_not} lists all the main 
lines observed for which we do not have atomic data for.
In most cases, lines have not been identified. Without
some estimates on their observed intensities, it is impossible to 
attempt some identifications. There are indeed many weaker lines in 
the CHIANTI spectrum which are not shown in the Table~\ref{tab:big_table}.
In several cases, however, observed lines are certain and their
identification suggested. Once atomic data are available for those lines,
we will be able to confirm those identifications.

\subsection{On the \ion{Ar}{x} line}

The \ion{Ar}{x} line deserves a special mention, given that 
it is potentially very important for elemental abundance measurements,
as discussed below.

There two lines, one from  \ion{Ar}{x} and one from \ion{Fe}{x}
which are predicted to be close in wavelength around 5537~\AA. 
In some literature observations, both lines are reported, but in others 
only one was, and sometimes the identification was not clear.

For both ions, we have atomic data rates that should be relatively accurate.
For \ion{Fe}{x} CHIANTI v.8 has atomic data from  \cite{delzanna_etal:12_fe_10},
while for \ion{Ar}{x} CHIANTI  has the scattering data 
of \cite{witthoeft_etal:07_f-like}. Both have been obtained within the UK
APAP network with the same $R$-matrix codes.
This means that the predicted intensities of these lines should be accurate.
The present model predicts that the \ion{Fe}{x} line should always
 be much weaker (by a factor of 5--10) than the  \ion{Ar}{x}   line,
regardless of the choice of abundances.
Also note that both lines are of class I.

The number of photons emitted by the \ion{Fe}{x} line
is predicted to be small, less than 1/10 than those of the 
 \ion{Fe}{x} 3454~\AA\ line (even considering photoexcitation which increases the ratio).  
The \cite{magnant-crifo:1973} study is one of the few where 
two lines of class I  are reported (although they were not identified as such), 
at 5533.4 and 5539.1~\AA.
In these observations, the \ion{Fe}{x} line
is indeed seldom observed. Its intensity does not always agree with 
prediction though. Also, the intensity of the nearby \ion{Ar}{x}   line
is  sometimes  higher, sometimes lower than the 
intensity of the  \ion{Fe}{x}   line, which is puzzling.
The other ground observations of the 1965 and 1966 eclipses do not 
report intensities for these lines, so we cannot cross-check the intensities
of these lines with other observations.

Regarding the wavelengths of these lines, 
we have cross-checked the \ion{Fe}{x} identifications, and a wavelength
of  5539~\AA\ is well constrained by other observations in the visible and 
UV (see \citealt{delzanna_etal:04_fe_10} for details but note that the level 
indexing in CHIANTI v.8 is different). This means that the wavelength in air
of this line should be 5538~\AA. 

On the other hand, the wavelength of the \ion{Ar}{x} is more uncertain, around 
5535~\AA\ (in air).
\cite{curtis_etal:1965} reports two lines, one at 5539.1~\AA\ of class I,
assigned to  \ion{Ar}{x}, and one unidentified  at 5533.4~\AA.
We believe that the  \ion{Ar}{x} identification is incorrect.

In Lyot's 1952 observation, only one line is reported, with 
an intensity even higher than that of  the \ion{Fe}{x} 3454~\AA\ line.
The measured wavelength by Lyot was 5539.5 (in air), so we  
originally thought that this was the  \ion{Fe}{x} line, although it cannot be.
On intensity grounds, this line can only be the  \ion{Ar}{x} line.
In their revised wavelengths,  \cite{aly_report} measured 5537,
which is however still quite different than the value of 5533.4 reported by
Curtis.

The \ion{Ar}{x} line was observed by \cite{koutchmy_etal:1974} to have 
the correct intensity ratio (5\%) with the green line, although it was
given  class III;  recent observations during the 2017 total eclipse confirm a 
stronger line around 5533.2~\AA\ (in agreement with Curtis's measurement), 
which must be the \ion{Ar}{x} line,
and a much weaker blend of two lines around 5538~\AA\ (S. Koutchmy, priv. comm.).
We therefore conclude that the \ion{Ar}{x} wavelength is 5533.4~\AA,
while the identification of the much weaker \ion{Fe}{x} line is still unclear.

\section{Estimates of line radiances off-limb}

As we have seen, we have provided as a rough guide 
estimated line radiances close to the solar limb. 
However, many interesting observations will be 
carried out at larger distances, so it is useful 
to know by how much line radiances fall off with the distance
from the Sun. 

This is not a trivial issue. One problem relates to the 
spatial distribution along the line of sight.  
We assume for our simple simulations a uniform distribution 
(i.e. spherical symmetry), which is only reasonable for the 
quiet Sun. In practice, we calculate the line emissivities
as a function of the radial distance, with the assumed 
parameters (densities, temperatures and abundances), out to 
10~\rsun. We then integrate these emissivities for each fixed distance
from the solar limb, assuming that for each radial distance 
the emissivities are the same (i.e. independent of longitude,
considering the equator).

Another problem is related to the photoexcitation. Again,
we assume uniform distribution of the disk radiances
 on the solar surface, and use the standard dilution 
factors.

However, the main variables are the radial distributions of the 
electron temperature and density. 
Previous modelling of a few important EUV coronal lines
carried out by \cite{andretta_etal:2012}
has shown significant differences in the predicted 
distribution of off-limb radiances, 
depending on which radial distributions one chooses.

The main problem is the radial distribution of the 
electron temperature, since there are very few 
(and contradictory) measurements. However, if the usual assumptions
of ionization equilibrium holds, a temperature can be obtained 
from the EM loci curves. That might not be the true electron temperature,
(indeed it is an ionisation temperature) 
but for the purpose of estimating the line radiances this is the best
approach. Once this is considered, it is usually found that,
at least up to 1.3\rsun, the temperature is constant. 
There are many results along these lines, for example from 
CDS \citep[see, e.g.][]{andretta_etal:2012} and SUMER 
 \cite[see, e.g.][]{landi_feldman:2003}. 
We have therefore assumed for our simulation a 
constant temperature of log $T$ [K]=6.14,  up to 1.5\rsun. 
After that, we have adopted 
the temperature distribution obtained by 
\cite{vasquez_etal:2003} with a semi-empirical method.

Fortunately, we have a better  understanding of the 
electron density, which is in any case the dominant parameter.
During the first two years of the SoHO mission, we obtained
a large number of off-limb observations and monitored the 
density using line ratios \citep{delzanna_thesis99,fludra99b}.
Near-simultaneous measurements of the polarized brightness
produced densities in reasonable agreement with those measured by CDS, 
for the quiet Sun \citep{gibson99a}.
We have therefore adopted the \cite{gibson99a} densities for our simulation.
Finally, we have chosen the photospheric abundances  of \cite{asplund_etal:09}.

\begin{figure}[htbp]
 \centerline{\includegraphics[width=7.0cm, angle=90]{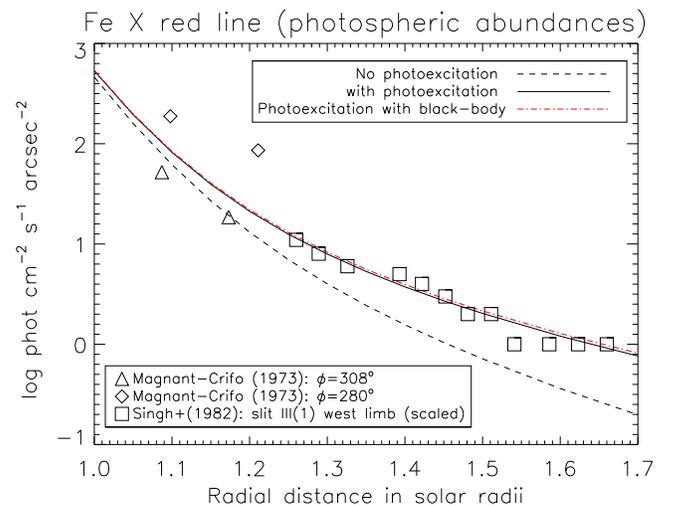}}
\caption{The radiance of the red line as a function of the radial distance,
for the quiet Sun.}
\label{fig:fe_10_red_offlimb}
\end{figure}

The results for the \ion{Fe}{x} red line are shown in Figure~\ref{fig:fe_10_red_offlimb}.
As it is well known, this line is strongly dependent on  photoexcitation.
For comparison, we are plotting the radiances measured by \cite{magnant-crifo:1973}
at two radial distances and position angles (in quiet regions). Some values are 
higher, but some are lower. We are however more interested to validate the
overall decrease with radial distance. 
We found an excellent 
eclipse observation of the red line by \cite{singh_etal:1982}, where one of the slit
positions (slit III (1) at the west limb) was more or less aligned with the 
radial direction, and the solar corona was quiet. The line intensities 
were not calibrated so we have scaled them and displayed them 
in  Figure~\ref{fig:fe_10_red_offlimb}. We can see very good
consistency in the decrease.

\begin{figure}[htbp]
 \centerline{\includegraphics[width=7.0cm, angle=90]{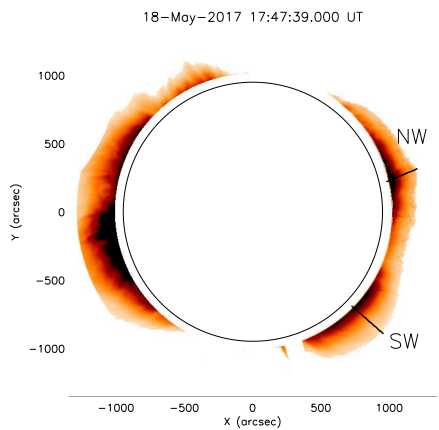}}
\caption{A negative  image of the \ion{Fe}{xiii} 10746~\AA\ infrared line, as observed by CoMP.}
\label{fig:comp}
\end{figure}

\begin{figure}[htbp]
 \centerline{\includegraphics[width=7.0cm, angle=90]{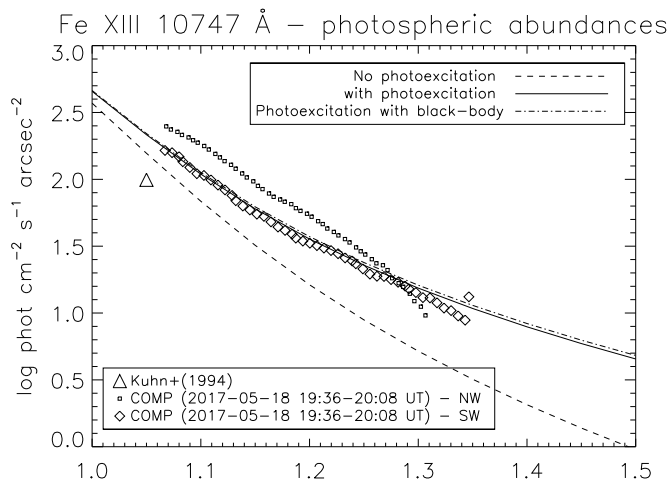}}
\caption{The off-limb decrease in the intensity of the 
\ion{Fe}{xiii} 10746~\AA\  infrared line 
 as a function of the radial distance, for the quiet Sun.}
\label{fig:fe_13_1_2}
\end{figure}

To further validate the model, we have considered the strongest of the 
 \ion{Fe}{xiii} infrared lines. 
As in the case of the red line, it is well known that the 
\ion{Fe}{xiii} infrared lines are very sensitive to photoexcitation
\citep[see, e.g.][]{young_etal:2003}.
To validate the overall decrease with distance, we have considered a 
CoMP observation when the Sun is relatively quiet, shown in 
 Figure~\ref{fig:comp}.
We have selected two radial directions, one in the SW, and one in the NW,
since the east limb was dominated by an active region. 
The SW had more signal and intensities could be measured up to 1.35\rsun.
We can see in  Figure~\ref{fig:fe_13_1_2} that the predicted decrease in the line 
intensities follows nicely the observed (scaled) one.

\section{A discussion on potential diagnostics using 
line ratios of the forbidden lines}

There is clearly a wide range of potential diagnostic applications
available to the visible and forbidden lines. 
For example, once the electron density is known, the intensity ratios
of lines from different ionisation stages of the same element can give
information on the ionisation temperature of the plasma. 
Below we just provide a few examples of line
intensity ratios which can be used to measure 
 the electron density and the relative  elemental abundances
using a few of  the strongest lines.

\subsection{Density diagnostics}

\begin{figure*}[htbp]
 \centerline{\includegraphics[width=6.cm, angle=90]{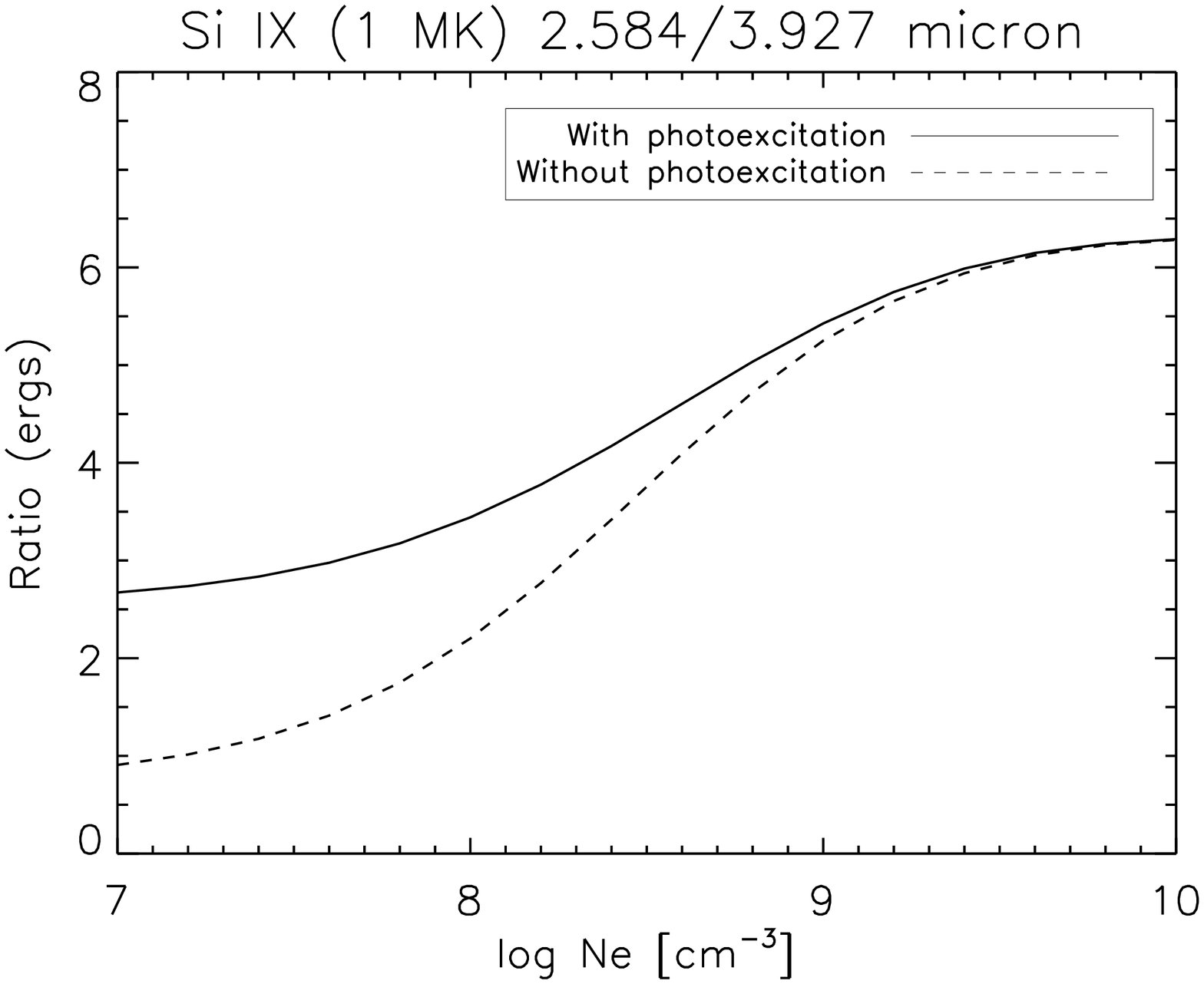}
\includegraphics[width=6.cm, angle=90]{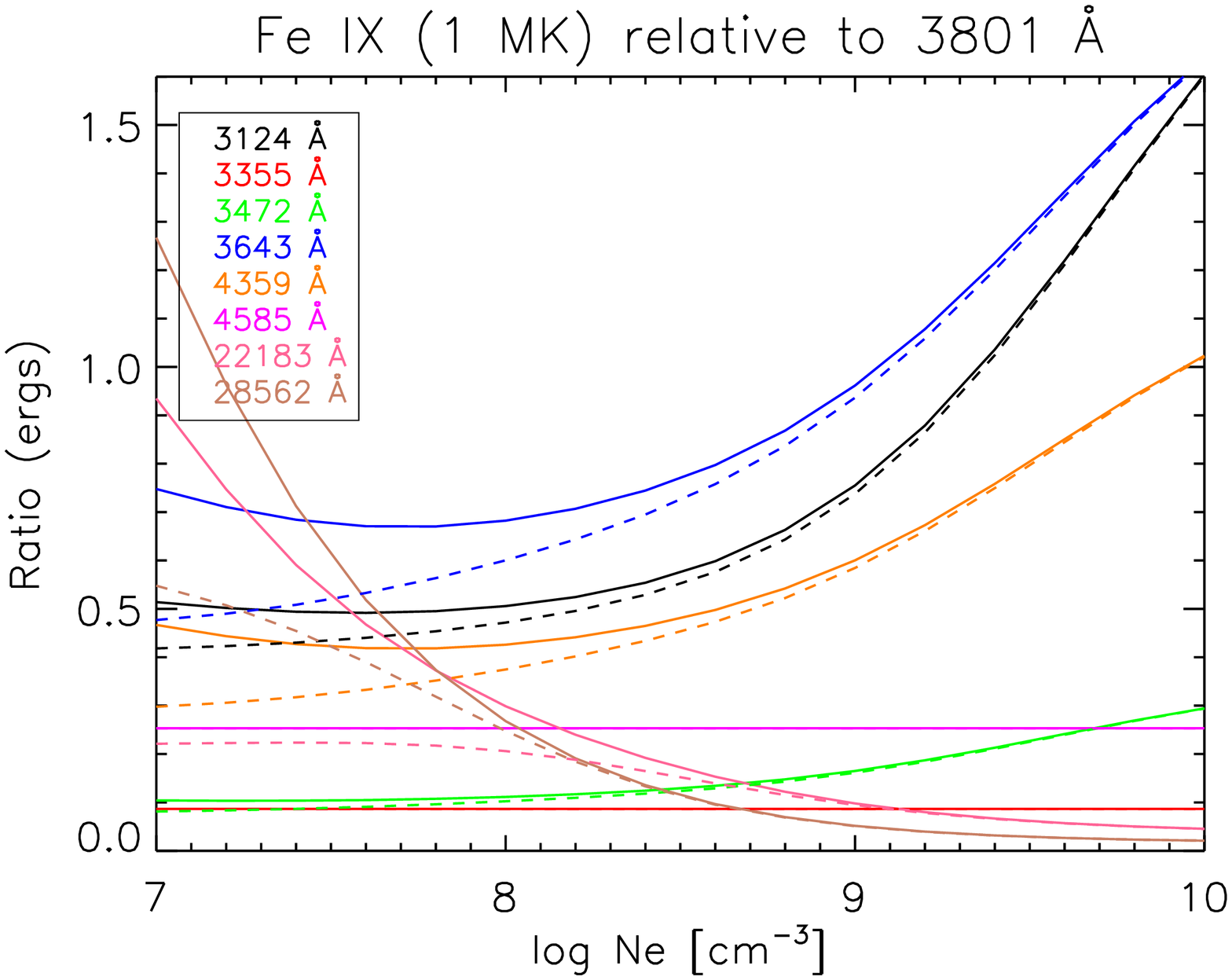}
}
 \centerline{\includegraphics[width=6.cm, angle=90]{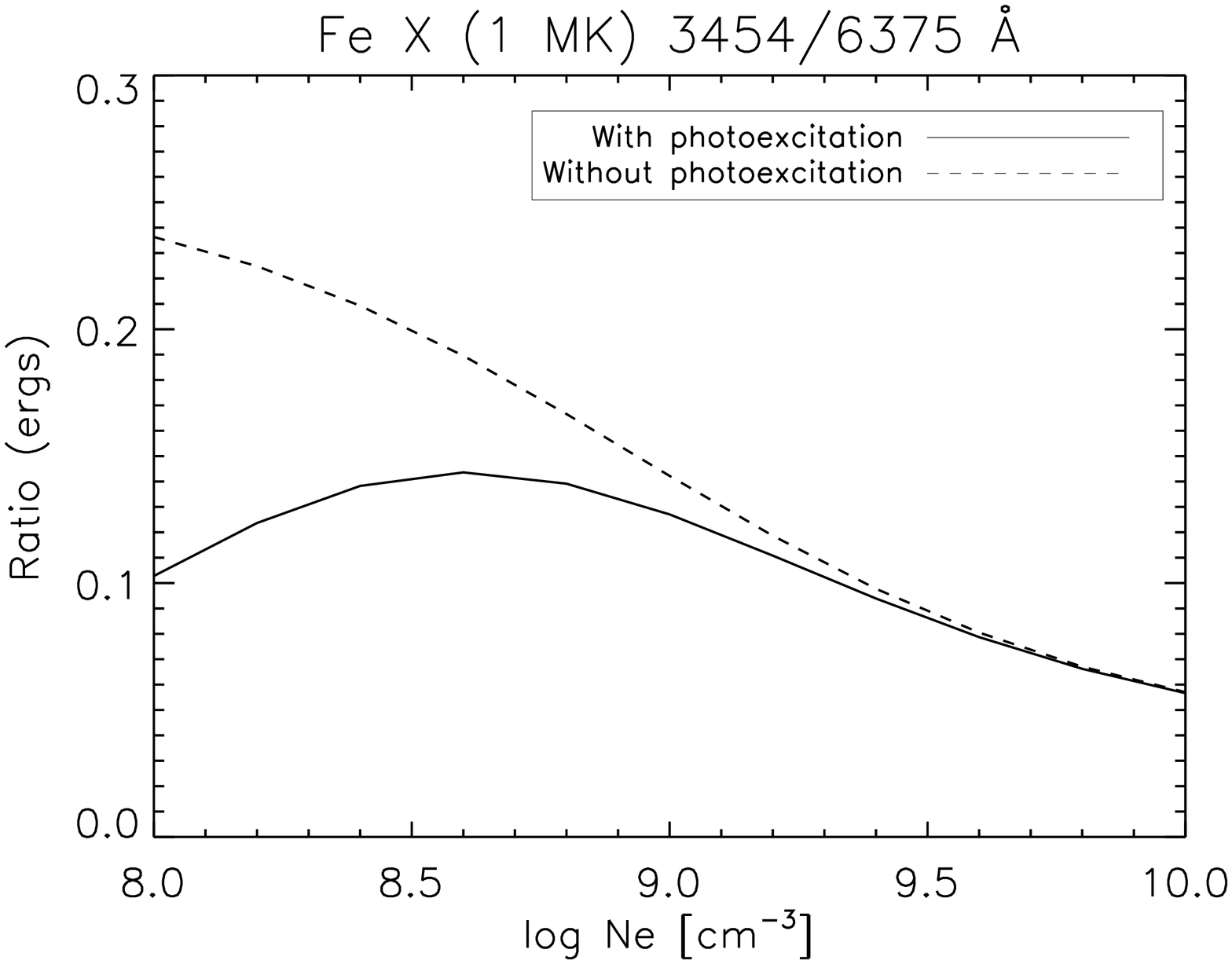}
\includegraphics[width=6.cm, angle=90]{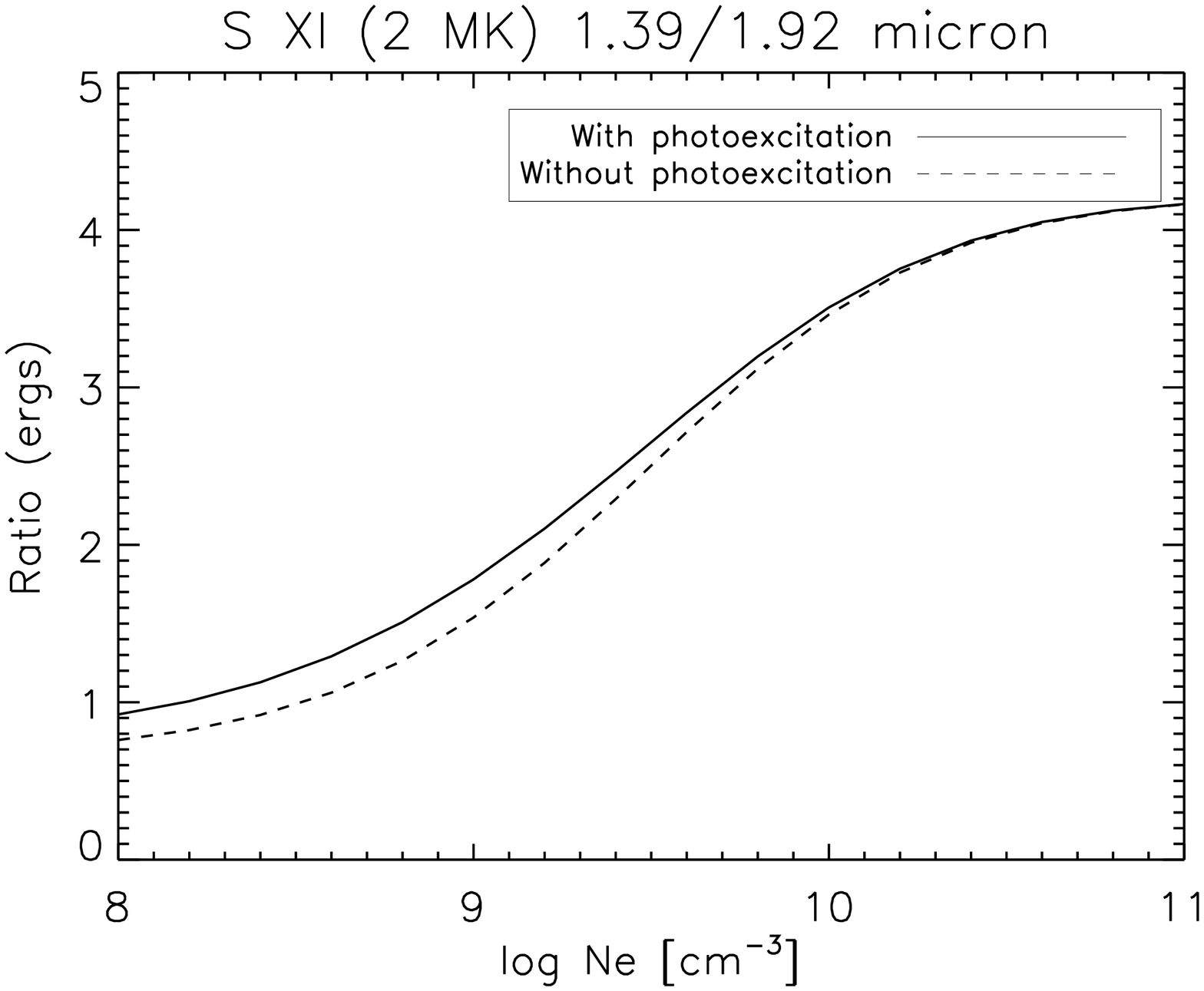}
}
 \centerline{\includegraphics[width=6.cm, angle=90]{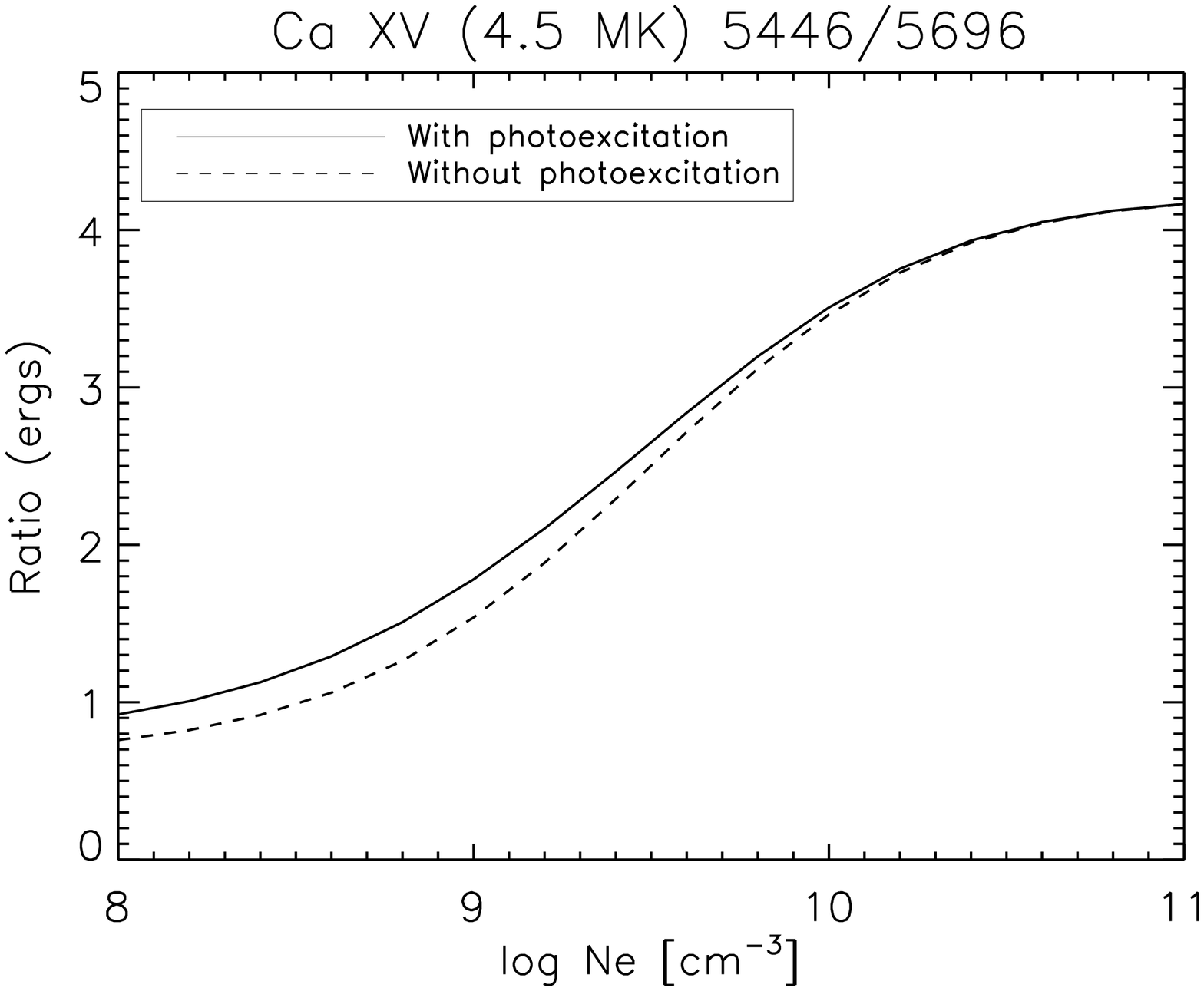}}
 \caption{Ratios of line radiances, without photoexcitation (full lines)
and with the present quiet Sun photoexcitation model included at 1.2\rsun 
(dashed lines).}
 \label{fig:densities}
\end{figure*}

All the visible and forbidden lines are strongly dependent on the 
electron density. However,  even if the absolute calibration 
(or the chemical abundances) are not known, there are several 
cases where densities can be obtained directly from line ratios.
In addition to the the well-known density diagnostics for active region plasma
from the \ion{Fe}{xiii} (infrared) and  \ion{Ca}{xv} lines,
there are several other diagnostics. 
A few examples  are provided here in Figure~\ref{fig:densities} and 
in  Table~\ref{tab:ne_table}.

An excellent diagnostic for the quiet Sun is provided by the \ion{Si}{ix} infrared lines,
considering that the lines are close in wavelength.
Another good diagnostic is the same ratio for the 
infrared  \ion{S}{xi} lines, which is sensitive to measure densities in active regions,
but should still be visible in the quiet Sun.

\subsection{Measuring the FIP effect}

\begin{figure}[htbp]
 \centerline{\includegraphics[width=7.0cm, angle=90]{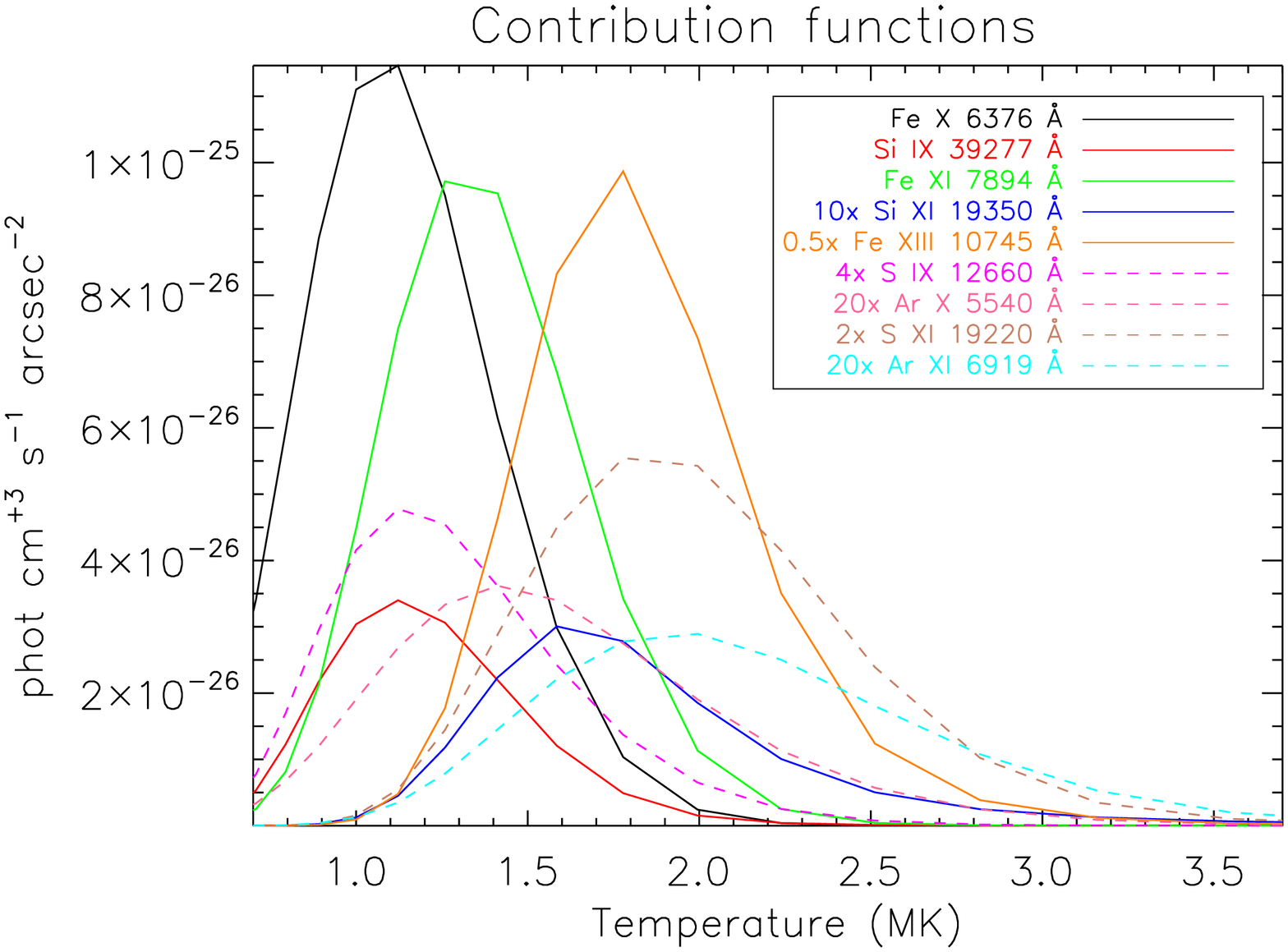}}
\caption{Contribution functions of a few of the main lines formed 
around 1-2 MK, calculated
with photospheric abundances. The full lines indicate ions from low-FIP
elements, while the dashed line those from high-FIP elements.
In a few cases, the values have been scaled as indicated.
}
 \label{fig:goft_qs}
\end{figure}

\begin{figure}[htbp]
 \centerline{\includegraphics[width=7.0cm, angle=90]{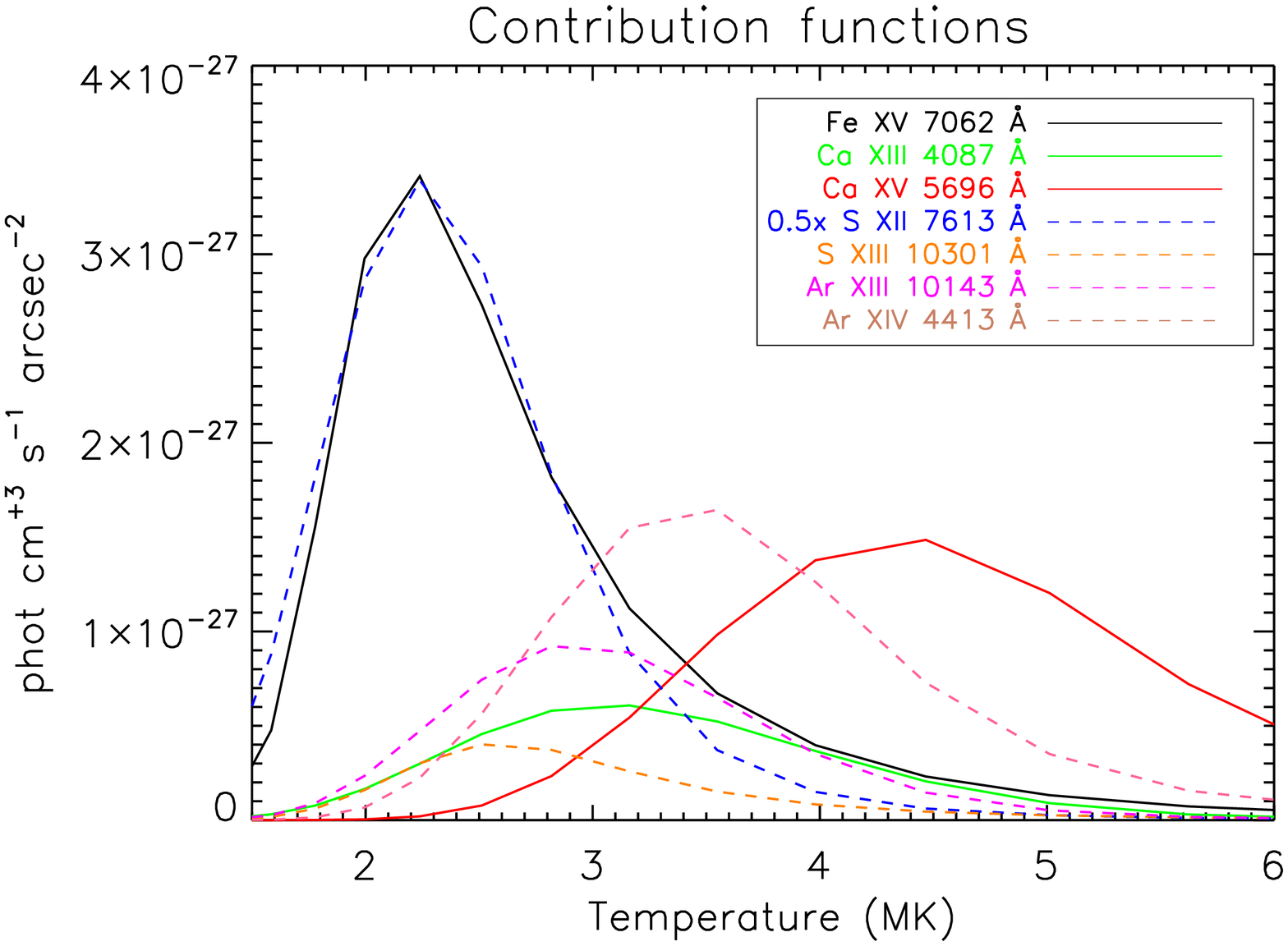}}
\caption{Contribution functions of a few of the main lines formed 
around 2--5 MK, calculated
with photospheric abundances. The full lines indicate ions from low-FIP
elements, while the dashed line those from high-FIP elements.
In a few cases, the values have been scaled as indicated.
}
 \label{fig:goft_ar}
\end{figure}

Also in this case there are plenty of diagnostics, depending on 
which wavelength region is considered, and the source region.
If one considers only the quiet Sun then most of the lines 
which are typically formed around 3 MK will not be observable,
and the list of potentially interesting lines reduces significantly.
 Figure~\ref{fig:goft_qs} shows the  contribution functions 
 of a few of the main lines formed 
around 1-2 MK (without photoexcitation included), 
calculated with photospheric abundances.
Within the visible, excellent diagnostics are available
with \ion{Ar}{x} and  \ion{Ar}{xi}, when observed in conjunction 
with e.g.  \ion{Fe}{x}, \ion{Fe}{xi}, and \ion{Fe}{xiii}.  
There are obviously many more lines from many other elements that 
will be observable by DKIST.

Many more diagnostics are available  when active regions are observed,
as it is clear from the list of lines in Table~\ref{tab:big_table}. 
In  Figure~\ref{fig:goft_ar} we show what we believe are the most important lines.
The best combination of lines for a 2 MK plasma is the 
ratio of the two strong  lines of \ion{S}{xii} at 7612~\AA\ and 
 \ion{Fe}{xv} at 7062~\AA, both of which have been observed regularly
at eclipses. 
The best diagnostic of all for the cores of active regions is the ratio
of \ion{Ar}{xiii} 10143~\AA\ with the  \ion{Ca}{xiii} 4087~\AA.
Their contribution functions are not very large, but in the cores of 
active region the emission measure at 3 MK is at least one or two 
orders of magnitude higher than the value at 1 MK, so these lines
will become very strong. Indeed the \ion{Ca}{xiii} has been observed to be strong
\citep[for a discussion of the Ca lines in the 1952 observation see ][]{mason:1975}.
On the other hand, the \ion{Ar}{xiii} 10143~\AA\ has not been observed yet
as far as we are aware.
Other important lines are those from \ion{S}{xiii}, \ion{Ar}{xiv} at 
4413~\AA\ and  potentially the  \ion{Ar}{xv} line if confirmed.
Several interesting possibilities are also shown in Table~\ref{tab:big_table}.
There are many transitions from K, Cl, Co, Mn that could potentially 
 allow abundance measurements in the future, if the lines will 
be observed and accurate  atomic data become available.

\subsection{The near infrared}

As already discussed, the near-infrared is basically 
an unexplored spectral region, with the exception of the 
few measurements we have cited.
It remains to be seen which spectral lines are observable.
According to our simulations, a number of weaker lines 
should be observable in the quiet Sun. Many of them were already pointed
out by  \cite{judge:1998}, but several were not.
We predict that, for example, the \ion{Fe}{xii} 2.2~$\mu$m
should be among the strongest lines. 
The  \ion{Al}{ix} at 2.2~$\mu$m should also be stronger than 
many other lines in this spectral region.
On the other hand, we predict very low signal in 
the transition-region lines such as \ion{Mg}{viii}, \ion{Si}{vii},
unless an active region is observed, or cooler material is present
along the line of sight.

The intensities of the strong  \ion{Si}{x} 1.43~$\mu$m reported by
\cite{munch_etal:1967} and  \cite{olsen_etal:1971} are in relatively good agreement
with our predicted radiances, while the \ion{S}{ix} 1.26~$\mu$m reported by 
\cite{olsen_etal:1971} is much stronger than any of our predictions. 
We know that the atomic model for \ion{S}{ix} needs improvement, and that 
on the basis of our large-scale calculations we expect an increase by 
a factor of 2--3, but that might not seem sufficient to explain the
observed intensity of this line.
 Regarding the other upper limits
listed by \cite{olsen_etal:1971}, they are compatible with our rough
estimates.

On a side note, we point out that several strong lines
from H and He are predicted to be visible, and indeed have been reported
by \cite{olsen_etal:1971}. 

As shown in  Figure~\ref{fig:densities}, the 
\ion{S}{xi} ratio is one of the best available 
across the whole visible/infrared spectrum,  considering that 
it is not much affected by  photoexcitation, and lines
are relatively close in wavelength. 
We expect these lines to be visible.

As Figure~\ref{fig:goft_qs} shows, excellent diagnostics 
for the FIP effect are available with \ion{S}{ix} 1.26~$\mu$m
and \ion{S}{xi} 1.92~$\mu$m, in combination with e.g. observations of 
the  \ion{Si}{ix} 3.93~$\mu$m and \ion{Si}{xi} 1.93~$\mu$m.
We note that three of these lines are in the Air-Spec spectral range, 
and appear to have been successfully observed in the 2017 eclipse.

\subsection{A few notes about DKIST CryoNIRSP}

In its current design, the CryoNIRSP will produce routine measurements
in a number of spectral bandpasses of about 20~\AA\ in the visible
and 100~\AA\ in the infrared \citep[see ][ for details]{fehlmann_etal:2016}.
In terms of coronal lines, the instrument will observe  the following lines:
\ion{Fe}{xiv} 5304~\AA\ and the  \ion{Fe}{xiii} doublet of infrared lines at 
10750, 10801~\AA, plus  several lines in the near-infrared, such as the 
\ion{S}{ix} at 1.25~$\mu$m,  \ion{Si}{x} 1.43~$\mu$m, the still unobserved
\ion{Fe}{ix} and \ion{Si}{ix} at 2.22 and 2.58~$\mu$m respectively, 
the  \ion{Mg}{viii} at 3.03~$\mu$m, and finally the still unobserved 
\ion{Si}{ix} at 3.93~$\mu$m.

The above selection of lines reflects 
the primary science goal of the DKIST coronal observations, to obtain 
magnetic field measurements. 
This set of lines has in fact an excellent  coverage
of the quiet Sun coronal emission around 1~MK. 
However, in principle the instrument is
capable of providing ground-breaking results in various other areas.  

For example, with its current settings, DKIST  will not observe the hot 
3~MK emission from active regions. 
As we have mentioned previously, there is evidence, at the highest
spatial resolution (Hi-C and AIA, 0.2 and 1\arcsec), that 
the 3~MK emission is not co-spatial with the 1--2~MK emission and 
has fundamentally different characteristics
\citep[see, e.g.][]{warren_etal:2008,delzanna:2013_multithermal}.

Considering the profound influence that 
active regions have on the overall restructuring and chemical 
composition of the lower corona, we strongly advise the addition of a few other channels
where at least a couple of strong lines formed around 3 MK are observed.
There are plenty of options, but well-known strong lines that have been 
observed are the  \ion{Fe}{xv} at 7062~\AA, the  \ion{Ca}{xiii} 4087~\AA,
and the famous yellow coronal line, the \ion{Ca}{xv} 5694~\AA. 

According to the model proposed by \cite{delzanna_etal:2011_outflows,bradshaw_etal:11},
 interchange reconnection
between the 3~MK loops and the surrounding corona 
continuously takes place high in the corona. This process is in principle capable
of injecting coronal plasma that was originally present in the closed
3 MK loops into the heliosphere. It is well known that this plasma 
has enriched low-FIP elements, so this is a possible way by which 
active regions enrich the low corona with low-FIP elements, as indeed 
has been observed in some instances with in-situ ACE measurements 
\citep{wang_etal:2009}.
According to the model, some of the 
 so-called coronal outflows, i.e. region connected to  
strong magnetic polarities (sunspots and moss) are  the signatures
of  outflows into the heliosphere. If this is the case, 
we need measurements of abundances, flows, densities, non-thermal widths
of 3~MK plasma, since these outflows are only seen in lines formed 
around 2--3 MK,  as shown by \cite{delzanna:08_flows}
and later confirmed by various authors \citep[see, e.g.][]{warren_etal:2011_outflows}.

To confirm this scenario, and in view of potentially important synergies
with the in-situ close-by observations from the Parker's Solar probe, and 
the in-situ plus  remote-sensing measurements  with the suite of instruments
aboard Solar Orbiter, it is of paramount importance to add to the DKIST 
CryoNIRSP the capability to measure abundances of 3~MK plasma. 
 Tracing the variations of the 
chemical composition from the low-corona to the solar wind is an important
science goal of Solar Orbiter, and DKIST could  provide invaluable information.

As briefly discussed in the previous 
section, the best combination of lines to measure the FIP effect are the 
\ion{Ar}{xiii} 10143~\AA\ with the  \ion{Ca}{xiii} 4087~\AA, together with the
\ion{S}{xiii}, \ion{Ar}{xiv}, if observed in combination to other lines
from low-FIP elements.
There is substantial evidence \citep[see, e.g.][]{delzanna:2013_multithermal} 
that in the low-corona 
the sulfur abundance  varies hand in hand with the argon abundance, in the 
cores of active regions.
However, there are in-situ measurements where this does not occur. 
Observing both argon and sulphur lines will hopefully shed light into 
the physical reasons as to why this happens. 
See the \textit{Living Review} by \cite{laming:2015} for more details.

\section{Summary and conclusions}

The present benchmark study is very satisfying for the visible 
forbidden lines, in that all the strongest lines classified 
with an `A' by  \cite{aly_report} are now firmly identified, and the latest
atomic data provide relatively good agreement with observations,
considering the various uncertainties in the calibration
of the spectra.
For a few spectral lines, we have provided new identifications.

Regarding the lines in class `B', most of them still await
identifications and/or the calculation of  atomic data.
Strangely enough, a few of the lines classified with a `C' 
have actually been observed in other eclipses and 
now have a firm identification.
We have also provided several tentative new identifications.

In the infrared region of the spectrum, we were unable
to provide any meaningful comparison, given the paucity of
observations. However, all the lines that have been 
observed are firmly identified, and we have corrected 
a few inaccuracies in the literature.
As already pointed out by \cite{judge:1998}, there are several 
coronal lines that should be observable in  this spectral region.
The advantage of observing in this region is the lower 
sky brightness and lower photospheric continuum emission.
The challenge is the transparency of the atmosphere to infrared
radiation. 

We hope that we have provided enough evidence that 
the visible and infrared forbidden lines can 
be used in the near future to measure chemical abundance
variations, electron temperatures and densities, alongside
coronal magnetic field measurements, if there is enough
signal in the polarisation measurements.
There are exciting opportunities ahead, to finally resolve some 
among the many mysteries of the solar corona using the 
 forbidden  visible and infrared  lines.

Regarding the atomic data, improvements in the 
energies and the atomic rates are needed for several ions. Some of them 
are in progress, but a significant amount of work still 
needs to be carried out before all the lines can reliably
be used for diagnostic purposes.
The issue of laboratory spectroscopy for astrophysics is
often overlooked by the funding bodies, and indeed 
the lack of reliable atomic data across 
the visible and infrared spectrum is no surprise.

The present rough guide relied on the atomic data available
within CHIANTI version 8 \citep{delzanna_chianti_v8}. 
The accuracy in the atomic data across
the entire wavelength range considered here varies considerably,
and further improvements are needed, as discussed in the Appendix. 
During the present study, we have highlighted several ions 
where no atomic data are yet available, an
issue that we will address in the near future.
Further discussions on specific ions of particular diagnostic
interest are left for a future paper.

\acknowledgements
GDZ  acknowledges support from  STFC (UK) and from SAO (USA)
during his visit to CfA.
EED was supported by NSF MRI: Development of An Airborne Infrared
Spectrometer (AIR-Spec) for Coronal Emission Line Observation, Award
Number: 1531549.

A special thanks goes to the Librarians at the CfA,
for providing access to various resources, in particular 
the Aly et al. (1962b) monograph.
The authors thank  V. Martinez-Pillet for pointing out the need 
for a preliminary  evaluation of the DKIST potential 
for coronal diagnostics. 

The authors are grateful to the following colleagues for useful 
comments and discussions:  H.E. Mason, L. Golub, P.R. Young,
 V. Andretta, D. Telloni, J. Raymond, 
K. Reardon, S. Koutchmy,  P. Judge., and  J. Samra

CoMP data are courtesy of the Mauna Loa Solar Observatory, 
operated by the High Altitude Observatory, 
as part of the National Center for Atmospheric Research (NCAR). 
NCAR is supported by the National Science Foundation.

CHIANTI is a collaborative project involving George Mason University, 
the University of Michigan (USA) and the University of Cambridge (UK).

\bibliographystyle{apj}
\bibliography{../../bib.bib}

\appendix

\section{A few notes about the atomic data}

The present assessment would not have been possible without
the accurate atomic cross sections for electron impact excitation
 recently produced by the UK APAP network
for the iron and nickel ions:
\ion{Fe}{viii}: \cite{delzanna_badnell:2014_fe_8};
\ion{Fe}{ix}:  \cite{delzanna_etal:2014_fe_9};
\ion{Fe}{x}: \cite{delzanna_etal:12_fe_10}; 
\ion{Fe}{xi}: \cite{delzanna_storey:2013_fe_11};
 \ion{Fe}{xii}: \cite{delzanna_etal:12_fe_12};
\ion{Fe}{xiii}: \cite{delzanna_storey:12_fe_13};
\ion{Fe}{xiv}:  \cite{liang_etal:10_fe_14};
\ion{Ni}{xi}: \cite{delzanna_etal_2014_ni_11};
\ion{Ni}{xii}: \cite{delzanna_badnell:2016_ni_12};
\ion{Ni}{xv}: \cite{delzanna_etal_2014_ni_15}.
For each of these ions extensive benchmarks have been carried out,
so the emissivities of the strongest lines should be reliable to 
within say a 10-30\%, depending on the line.
Within the CHIANTI version 8 database, such level of accuracy is not
present for most of the other ions in the visible and infrared.
However, for most the lines of interest, the present atomic data
should be accurate enough for the present rough guide.

For \ion{Fe}{xv} the calculations of  \cite{berrington_etal:05_fe_15}
have a similar level of accuracy, while for the 
\ion{Ni}{xiii} and \ion{Ni}{xvi}, the calculations of 
\cite{bhatia_doschek:1998} and 
\cite{bhatia_doschek:1999} will need some improvement. 

Considering the magnesium, sulphur, and  silicon ions, the 
cross sections for the B-like \ion{Mg}{viii}, \ion{S}{xii},  
and \ion{Si}{x} 
 calculated by \cite{liang_etal:2012} should be accurate,
while those for the other ions need improvement, because of the 
limited calculations 
(\ion{Si}{ix}, \ion{S}{ix} by \citealt{bhatia_landi:2003})
or because data were interpolated 
(\ion{S}{xi} 
and \ion{S}{xiii}).  

 Considering the calcium and argon ions, the cross sections  
for the F-like 
\ion{Ca}{xii} and \ion{Ar}{x} calculated by \cite{witthoeft_etal:07_f-like}
should be relatively accurate, while those for the O-like  
\ion{Ca}{xiii} and \ion{Ar}{xi} by \cite{landi_bhatia:2005} and 
\cite{landi_bhatia:2006} should be improved. The cross-sections for
C-like \ion{Ca}{xv} calculated by \cite{aggarwal_etal:1991}
should be relatively accurate, while those of 
\ion{Ar}{xiii} \citep{dere_etal:1979} need improvement. 
The data for the B-like 
\ion{Ar}{xiv} calculated by \cite{liang_etal:2012} should be accurate.

The atomic data for some of the above ions have been calculated and 
will be made available with the next version of CHIANTI. 
We also note that the UK APAP network will be carrying out 
new large-scale calculations for the C-, N-, and O-like isoelectronic
sequences, where the current atomic data are not consistently
accurate. This will improve considerably the atomic data 
for several ions considered in the present paper.

\end{document}